\documentclass[fleqn,usenatbib]{mnras}

\usepackage{newtxtext,newtxmath}
\usepackage[T1]{fontenc}
\usepackage{ae,aecompl}

\usepackage{graphicx}	% Including figure files
\usepackage{amsmath}	% Advanced maths commands
\usepackage{ulem}

\newcommand{\Msun}{\,{\rm M}_{\sun}}

\newcommand{\logm}{\,{\Delta \log{M}}}
\newcommand{\Gyr}{\,{\rm Gyr}}
\newcommand{\kpc}{\,{\rm kpc}}
\newcommand{\kms}{\,{\rm km s^{-1}}}

\title[Jeans Modelling]{What to expect when using globular clusters as tracers of the total mass distribution in Milky Way-mass galaxies}

\author[M. E. Hughes et al ]{Meghan E. Hughes$^{1}$\thanks{E-mail: M.Hughes1@2013.ljmu.ac.uk},
Prashin Jethwa$^{2}$, 
Michael Hilker$^{3}$, 
Glenn van de Ven$^{2}$, 
\newauthor
Marie Martig$^{1}$,
Joel L. Pfeffer$^{1}$,
Nate Bastian$^{1}$,
J.M. Diederik Kruijssen$^{4}$, 
\newauthor
Sebastian Trujillo-Gomez$^{4}$,
Marta Reina-Campos$^{4}$,
Robert A. Crain$^{1}$
\\
$^{1}$Astrophysics Research Institute, Liverpool John Moores University, 146 Brownlow Hill, Liverpool L3 5RF, UK\\
$^{2}$ Department of Astrophysics,  T\"{u}rkenschanzstra{\ss}e 17, 1180 Vienna, Austria\\
$^{3}$European Southern Observatory, Karl-Schwarzschild-Stra{\ss}e 2, 85748, Garching, Germany\\
$^{4}$Astronomisches Rechen-Institut, Zentrum f\"{u}r Astronomie der Universit\"{a}t Heidelberg, M\"{o}nchhofstra{\ss}e 12-14, 69120 Heidelberg, Germany\\
}

\date{Accepted 2021 January 20. Received 2021 January 20; in original form 2020 October 2}

\pubyear{2021}

\begin{document}
\label{firstpage}
\pagerange{\pageref{firstpage}--\pageref{lastpage}}
\maketitle

\begin{abstract}
Dynamical models allow us to connect the motion of a set of tracers to the underlying gravitational potential, and thus to the total (luminous and dark) matter distribution. They are particularly useful for understanding the mass and spatial distribution of dark matter (DM) in a galaxy. Globular clusters (GCs) are an ideal tracer population in dynamical models, since they are bright and can be found far out into the halo of galaxies. We aim to test how well Jeans-Anisotropic-MGE (JAM) models using GCs (positions and line-of-sight velocities) as tracers can constrain the mass and radial distribution of DM halos. For this, we use the E-MOSAICS suite of 25 zoom-in simulations of L* galaxies. We find that the DM halo properties are reasonably well recovered by the JAM models. There is, however, a strong correlation between how well we recover the mass and the radial distribution of the DM and the number of GCs in the galaxy: the constraints get exponentially worse with fewer GCs, and at least 150 GCs are needed in order to guarantee that the JAM model will perform well. We find that while the data quality (uncertainty on the radial velocities) can be important, the number of GCs is the dominant factor in terms of the accuracy and precision of the measurements. This work shows promising results for these models to be used in extragalactic systems with a sample of more than 150 GCs.

\end{abstract}


%%%%%%%%%%%%%%%%%%%%%%%%%%%%%%%%%%%%%%%%%%%%%%%%%%

%%%%%%%%%%%%%%%%% BODY OF PAPER %%%%%%%%%%%%%%%%%%

\section{Introduction\label{1}}
The distribution of mass within a galaxy contains information about its formation and evolution. It also helps us to understand how the dark and baryonic matter are linked (e.g. \citealt{Blumenthal1986,Remus2017}). Mass distributions may be obtained via a variety of techniques such as strong gravitational lensing (e.g. \citealt{Auger2010,Sonnenfeld2013}), the virial theorem (e.g. \citealt{Watkins2019}) and dynamical modelling (e.g. \citealt{Tortora2014,Poci2017}).Dynamical models are used to connect the motion of a set of tracers to the gravitational potential. This allows kinematic data sets to be turned into information  not only about the distribution of mass within a galaxy (including the DM content) but also the intrinsic shape (e.g. \citealt{vandenBosch2008}), the stellar initial mass function of its composite stellar population (e.g. \citealt{Thomas2011,Posacki2015,Tortora2016,Li2017}), the baryonic to dark-matter mass ratio (e.g. \citealt{Thomas2011,Zhu2016_elliptical}), the merger history of the galaxy (e.g. \citealt{Schulze2020}) and the intrinsic properties of the tracer population (e.g. \citealt{Schuberth2010}).\par

Jeans models \citep{Jeans1915,Binney1980,Merritt1985,Dejonghe1992} involve solving the Jeans equations (spherical, axisymmetric or triaxial) for the kinematics of a galaxy based on a parameterisation of the galaxy mass distribution. Jeans models are commonly used to provide initial insights for computationally slower (but more sophisticated) models, such as Schwarzschild models \citep{Schwarzchild1979} and made-to-measure models \citep{Syer1996}. They also have the advantage that no functional form has to be assumed for the underlying distribution functions. Jeans models have been used to determine the distribution of the DM of all kinds of galaxies, from dwarfs (e.g. \citealt{Kleyna2001,Battaglia2008,Strigari2008,Lokas2009,Walker2009}) to ellipticals (e.g. \citealt{Napolitano2009,Schuberth2010,Deason2012,Agnello2014}). In early-type galaxies the GC velocities obtained by the SLUGGS survey (e.g. \citealt{Forbes2017}) has been combined IFU kinematic maps from the ATLAS$^{\mathrm{3D}}$ survey to model the total-mass profiles of a sample of 21 galaxies in the stellar mass range $10^{10} < M_{*}/M_{\odot}<10^{11.6}$ \citep{Bellstedt2018}. \citet{Bellstedt2018} find that the mass and density profile slope measured through the Jeans modelling are consistent with those measured in the inner regions of galaxies by other studies and using other techniques.

\citet{Leung2018} combined stellar kinematics from CALIFA \citep{Sanchez2016} with gas kinematics from the EDGE \citep{Bolatto2017} survey and found that the Jeans models, along with the Shwarzchild models and Asymetric Drift Correction can recover the dyamical mass within 1Re to within 20 per cent, but cautioned that assumptions may break down in the inner regions. In addition \citet{Scott2015} used Jeans models to calculate the dynamical masses of 106 SAMI \citep{Croom2012} galaxies. Jeans models have also been used in the ATLAS$^{\mathrm{3D}}$ \citep{Capellari2011} and the MaNGA \citep{Bundy2015} surveys to find variations in the stellar initial mass functions of early type and late type galaxies \citep{Cappellari2012,LiH2017}.

\par

In the Jeans-Anisotropic-MGE (JAM) modelling technique \citep{Cappellari2008,Cappellari2012,D'Souza2013,Watkins2013}, the potential and tracer densities are given as inputs in the form of multi-gaussian-expansions (MGE) \citep{Emsellem1994,Cappellari2002}. \cite{Watkins2013} extended the JAM model by removing the need for binning and working directly with the discrete data. Fitting each tracer particle individually means that quality cuts in the data are no longer needed and the likelihoods can be extended to easily incorporate further information such as the metal abundances. \cite{Watkins2013} applied these models to resolved stars in the GC $\omega$\,Centauri to find the velocity anisotropy, the inclination angle, a $V$-band mass-to-light ratio and a distance that are all in agreement with the values found in previous studies. Based on the dynamical models of \citet{Watkins2013}, \citet{denBrok2014} constructed dynamical models of the GC M15, again using the discrete fitting method. They were able to show that the models reproduced the radial variation of the mass-to-light ratio found in other studies and theoretical predictions.\par

\citet{Zhu2016_dwarf} extended the models by \citet{Watkins2013} to include multiple populations in a new chemo-dynamical axisymmetric Jeans model. They applied this model to several mock data sets for the dwarf spheroidal galaxy Sculptor, where they considered different stellar populations tracing the same potential. Where most Jeans modelling techniques compute a likelihood in the kinematics, in this case a combined likelihood in position, metallicity and kinematics is used to constrain the mass profile, velocity anisotropy and internal rotation of the dwarf galaxy. This type of model allowed \citet{Zhu2016_dwarf} to show that stars in Sculptor naturally separated into two populations - metal rich and metal poor. The two populations have different spatial distributions, velocity dispersions and rotation. \citet{Zhu2016_elliptical} further extended the axisymmetric Jeans model to include three dynamical tracer populations and also fit the integrated light stellar kinematic data in the inner region of the giant elliptical galaxy NGC 5846. The three dynamical tracer populations were the planetary nebulae (PNe) and two GC subpopulations. Using this method they constrained the mass distribution including the DM fraction and the internal dynamics of each tracer population.\par

It is clear that dynamical modelling techniques can reveal a lot about a stellar or galactic system.  It is therefore important to test these kinds of models on hydrodynamical simulations to fully understand what biases may be present when they are applied to real systems. These biases may be driven by the sample size of the dynamical tracers, the data quality or the intrinsic properties of the stellar or galactic system being modelled.  \par

JAM has been extensively tested on >1000 simulated galaxies by \citet{Li2016}. They used the Illustris project \citep{Genel2014,Vogelsberger2014,Nelson2015} to select massive galaxies and constructed a dynamical model for each galaxy. In this study, \citet{Li2016} construct kinematic maps and brightness maps of the galaxies, which are then used as inputs for the JAM modelling. They find that the total mass enclosed within 2.5 $R_{e}$ is constrained to within 10 per cent. They also find that the 1$\sigma$ scatter in the recovered stellar mass-to-light ratio $M^*/L$ is 30-40 per cent of the true value and this accuracy depends on the triaxial shape of the galaxy. \par

Similarly, \citet{El-Badry2017} used simulations from the FIRE project to test the reliability of Jeans models on low-mass galaxies. \citet{El-Badry2017} use the stellar
radial velocity profile and number density profile as inputs into their Jeans models. They connect the results from the Jeans models to the gas inflow and outflow of these low-mass galaxies and find that the Jeans model overestimates a galaxy's dynamical mass during periods of post-starburst gas outflow and underestimates it during periods of net inflow. They place a lower limit of $20$ per cent uncertainty in the mass measurements of gas-rich galaxies, but this is reduced to $10 \%$ in gas-free galaxies.\par

The \citet{Li2016} study is based on the assumption that we have a kinematic map for a full galaxy, and the study of \citet{El-Badry2017} is based on the assumption that we can extract a radial velocity profile of the stars. However, the dynamics of the full galaxy halo must be used to obtain as much information as possible about the properties and formation of a galaxy. This includes the very outskirts of the galaxy, where the most information about the merger history is contained. With the advent of large surveys (e.g. SLUGGS, Fornax 3D, \citealt{Sarzi2018,Fahrion2020})it is an opportune time to take full advantage of the GC  and planetary nebulae (PNe) survey data. GCs and PNe are bright tracers in distant galaxies that probe far out into a galaxy's halo and therefore are particularly useful for constraining the radial distribution and mass of the DM halo \citep{Schuberth2010} and the merger history of the galaxy \citep{Schulze2020,Trujillo2020}.\par

The suite of E-MOSAICS simulations \citep{Pfeffer2018,Kruijssen2019} forms and evolves GCs fully self consistently alongside their host galaxies. This gives us the unique opportunity of using these simulated GCs as tracers in the JAM model to test our ability to recover the mass profile of a galaxy using just the line-of-sight velocities of  the GCs. We test this method on 25 zoom-in simulations of MW-mass galaxies and their associated GC populations.\par

This paper is organised as follows, Section \ref{2} outlines the relevant details of the simulations. In Section \ref{3} we describe the Jeans model used for this work and the inputs. Section \ref{4} contains the outputs of the model and our first steps to interpret them, including the enclosed mass profiles. In Section \ref{5} we discuss how the properties of the GC system, such as the spatial distribution, number and line-of-sight velocity error may affect the recovery of the DM mass profiles. In Section \ref{7} we correlate the DM profile recovery with other galaxy properties and discuss the resulting correlations and finally we summarise and conclude in Section \ref{8}.

\section{Simulations \label{2}}
\begin{center}
\begin{table}
 \caption{\label{table:GCnumber} The number of GCs in each of the 25 simulated galaxies for 3 different GC parameter restrictions. Note that the radius cut is based on projected x and y coordinates after the galaxy has been aligned as edge-on.}
\begin{tabular}{ |c|c |c|c| } 
 \hline
 Simulation &All GCs& Age>8 Gyr & Age > 8 Gyr , R > 2kpc  \\
 \hline
MW00 &252&245& 101 \\
MW01&642&382& 186 \\
MW02 &841&817& 400 \\
MW03 &547&534& 206\\
MW04 &264&251& 99\\
MW05 &951&949& 340\\
MW06 &441&328& 125\\
MW07 &251&117& 58\\
MW08 &200&75& 18\\
MW09 &255&178& 93\\
MW10 &1012&494& 295\\
MW11 &205&134& 66\\
MW12 &1013&810& 394\\
MW13 &280&168& 109\\
MW14 &827&239& 179\\
MW15 &551&37& 30\\
MW16 &504&442& 341\\
MW17 &337&108& 97\\
MW18 &121&61& 51\\
MW19 &108&73& 31\\
MW20 &385&137& 59\\
MW21 &181&146& 122\\
MW22 &365&252& 200\\
MW23 &711&395& 241\\
MW24 &340&102& 77\\
 \hline
 \end{tabular}
 \end{table}
\end{center}
This work uses the E-MOSAICS (MOdelling Star cluster population Assembly In Cosmological Simulations within EAGLE) suite of zoom-in simulations \citep{Pfeffer2018,Kruijssen2019}.  We use the volume-limited sample of 25 MW-mass galaxies from the high-resolution 25 cMpc volume EAGLE simulation (Recal-L025N0752; \citealt{Schaye2015}). The sample is chosen solely based on a total mass cut of $7 \times 10^{11} < M_{200}/M_{\odot} < 3 \times 10^{12}$ and therefore probes a variety of formation histories and is a representative sample of MW-mass galaxies. This, alongside being the largest set of hydrodynamical simulations that simultaneously follow the co-formation and evolution of galaxies and their GC populations in a cosmological context makes the E-MOSAICS simulations an ideal tool to test the use of GCs as tracers for Jeans modelling. We outline the relevant details of the simulations here, but refer the interested reader to \cite{Pfeffer2018} and \citet{Kruijssen2019} for detailed descriptions. \par

E-MOSAICS couples the MOSAICS subgrid model \citep{Kruijssen2011,Pfeffer2018} for star cluster formation and evolution to the EAGLE (Evolution and Assembly of GaLaxies and their Environments) galaxy formation and evolution model \citep{Schaye2015, Crain2015}. Each stellar particle, when formed from a gas particle, can form a number of subgrid star clusters depending on its local cluster formation efficiency \citep{Kruijssen2012} and mass function, with initial cluster masses stochastically sampled from a Schechter initial cluster mass function \citep{Schechter1976} with an environmentally dependent truncation mass \citep{ReinaCampos2017}. This star cluster population then inherits the age, chemical composition, position and velocity of its parent stellar particle. The star clusters formed in the MOSAICS model reproduce the observed properties of young clusters in the nearby Universe \citep{Pfeffer2019YC}.\par

The evolution of the star clusters is then followed alongside their host galaxy. Star cluster mass loss occurs in the form of stellar evolution (tracked by the EAGLE model, \citealt{Wiersma2009}), tidal shocks and two-body relaxation \citep{Kruijssen2011}. Complete disruption of clusters via dynamical friction is applied in post-processing.  \par

The E-MOSAICS simulations have been shown to reproduce many key observables of young and old cluster populations. The simulations have proven that GCs are powerful tracers of both the formation and evolution and general properties of Milky Way-mass galaxies. We refer to \cite{Kruijssen2019} for an overview and give a brief set of examples here. The age-metallicity-kinematic distribution of GCs has been shown to be broadly consistent with that of the Milky Way \citep{Kruijssen2019b,Kruijssen2020,Pfeffer2020,Trujillo2020}. The simulations have been used to investigate the disruption of GCs and have shown that the fraction of field stars in both the bulge and halo of the simulated galaxies is similar to that of the Milky Way \citep{Reina-Campos2018,Hughes2020,Reina-Campos2020}. The simulations have also shown their ability to use GCs to predict properties of dwarf galaxies that now cause stellar streams \citep{Hughes2019} and have been shown to reproduce the GC colour-magnitude relation \citep{Usher2018}. Another observable reproduced by the E-MOSAICS simulations is the relation between the number of GCs and the host galaxy mass \citep{Kruijssen2019,Bastian2020}. In addition, the simulations have made predictions for future observables regarding the formation environments and initial demographics of GCs, to be tested with JWST and the thirty-metre class telescopes \citep{Reina-Campos2019FH,Pfeffer2019P,Keller2020}. Taken together, these results make E-MOSAICS the best currently available suite to achieve the goals of this paper.\par

For the majority of our analysis, we use a fiducial sample of GCs defined as all star clusters with a present ($z=0$) mass $>10^5 \Msun$, old ages ($>8 \Gyr$), and excluding the innermost clusters (projected $R > 2 \kpc$). We make the mass cut to be consistent with what is likely to be observable (in external galaxies) with current telescope facilities. In addition we note that although the high mass end of the simulated GC mass function is in good agreement with that of the MW and M31, E-MOSAICS produces too many low mass clusters (likely due to under disruption, \citealt{Pfeffer2018,Kruijssen2019}). Thus, the mass cut also works towards mitigating this effect. We make the age cut to be consistent with what is considered a `traditional' GC, i.e. an old and massive bound star cluster. Also, this age cut will remove young disc clusters, which are often excised from spectroscopic studies because of extinction in the disc. Finally, we make the inner radius cut since this is the region where most observational studies are likely to be affected by crowding (although see e.g. \citealt{Fahrion2020} for an example of GCs being identified and velocities extracted in the inner parts of a galaxy using MUSE), therefore making it difficult to identify and get reliable velocity measurements for GCs. 
In addition to our fiducial sample, we also test the effect of not including cuts on the age and the radius. However, overall the results are not systematically affected by these selections.
For reference, we show the number of GCs in each simulated galaxy, for each selection in Table \ref{table:GCnumber}, keeping in mind that the clusters are always restricted to be more massive than $10^5 \Msun$.\par

\section{The Jeans Model \label{3}}
\subsection{Introducing the coordinate system\label{3.1}}
For this work we use the cylindrical version of the Jeans equations and therefore work in cylindrical polar coordinates ($R$, $\theta$, $z$). However for most of the equations presented here we work in projected Cartesian coordinates, where ($x',y',z'$) represent the projected coordinates on the plane of the sky. The $x'$-axis is aligned with the galaxy's projected major axis, the $y'$-axis with the projected minor axis and the $z'$-axis lies along the line-of-sight such that the vector is positive in the direction away from the observer. We also perform our calculations using the line-of-sight velocity ($v_{\mathrm{z'}}$) such that the vector is also positive in the direction away from the observer. The $v_{\mathrm{z'}}$ is calculated by subtracting the mean line-of-sight velocity of the galaxy from that of the star particle. \par

\subsection{Jeans-Anisotropic-MGE\label{3.2}}
\begin{figure*}
	\includegraphics[width=\linewidth]{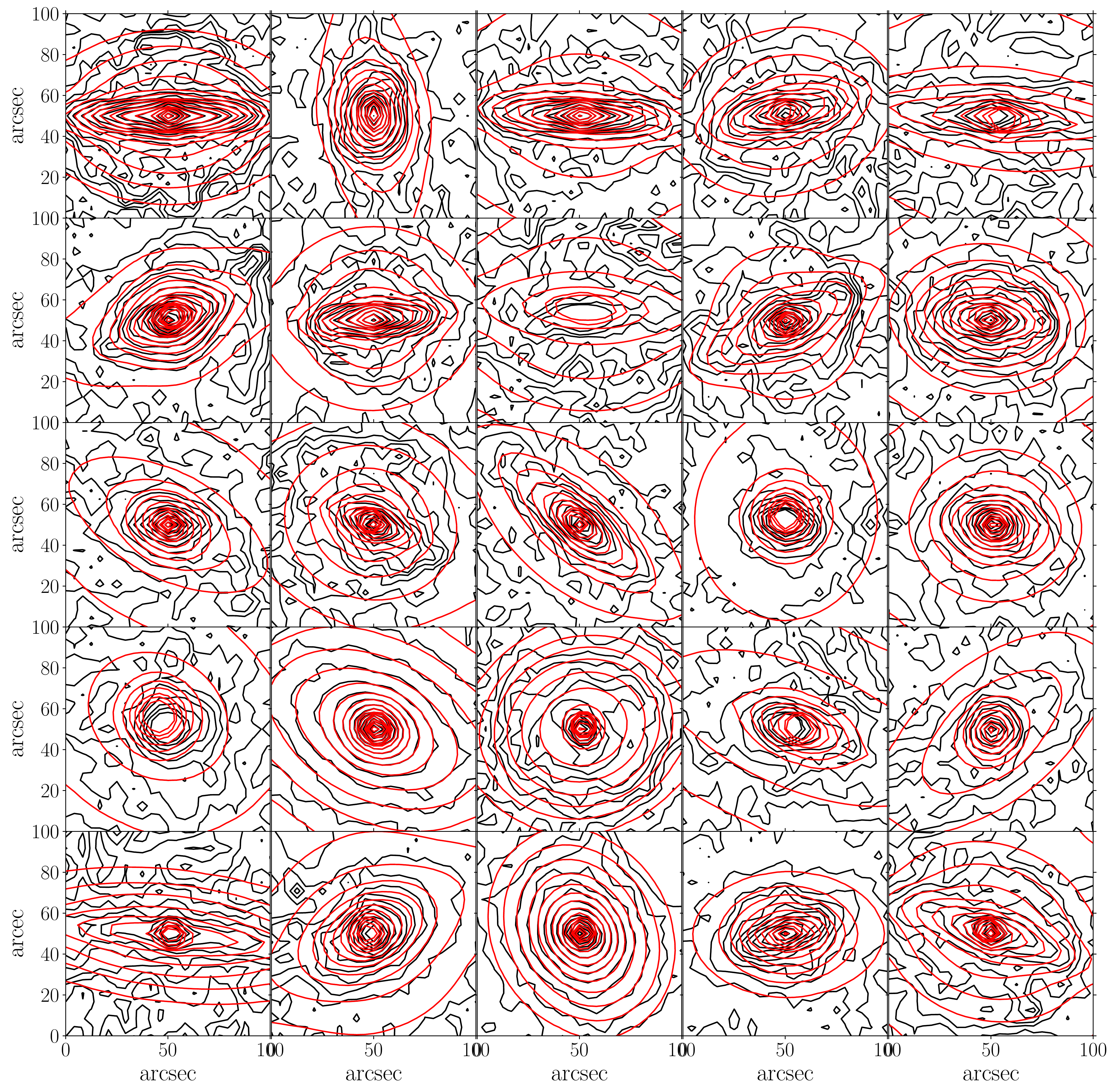}
    \caption{MGE fits for all 25 galaxies. The black contours show the distribution of baryons in the galaxies and the red ellipses show the MGE fits to the simulations. The galaxies are ordered from left to right and then top to bottom in increasing number, with MW00 in the top left and MW24 in the bottom right. The galaxies are projected at a distance of 1Mpc and we show the inner 100 arcsec square (100 arcsec equates to $\approx 480$ pc).}
    \label{fig:allgal_contours.pdf}
\end{figure*}

To determine the DM halo density distribution we use a Jeans model with parameterised potential and tracer densities and maximise the likelihood of each parameter by comparing the velocity outputs with the measured line-of-sight velocity at multiple positions in the galaxy. In conjunction with the Jeans theorem, we assume a steady-state, time-independent form of the potential for each galaxy.\par

We model the simulated data-set using the extended version of the axisymmetric Jeans Anisotropic MGE formalism. This particular fomalism takes the potential and tracer densities in the form of a multi-gaussian-expansion \citep{Emsellem1994,Cappellari2002}. The MGE is a series of 2D gaussians that provide information about the shape and intensity of a distribution. The MGE method developed by \citet{Emsellem1994} (based on \citealt{Monnet1992}) has the benefit of being able to perform deprojection analytically and efficiently. We use the \citet{Watkins2013} implementation of the JAM formalism as it removes the need to spatially bin data. This means we can directly pass in the line-of-sight velocities and positions of each GC as a discrete data point.\par

The goal of this work is to determine how well the JAM model recovers the radial distribution and the mass of the DM halo when we use GCs as our tracer population.  We use the JAM code made publicly available by \citet{Watkins2013}\footnote{https://github.com/lauralwatkins/cjam.git} to calculate the first- and second-moments of the line-of-sight velocity. The code requires as inputs:
\begin{itemize}
\item The tracer density, characterised as an MGE. In this case we are using GCs as tracers and we assume a spherical distribution for the GC population (i.e. we assume the system is not flattened in any direction). Therefore we fit a 1D MGE to the projected $r' = \sqrt{x'^2 + y'^2}$ GC positions. We investigate the effect of assuming a spherical distribution in Section \ref{5}.
\item The galaxy density, characterised as an MGE. We split our galaxy potential into two components. Firstly, the baryonic component, for which we use the MGE FIT SECTORS software (which is on the python package index,\citealt{Cappellari2002}) to fit a 2D axisymmetric MGE directly to the $x'$ and $y'$ positions of the star and gas particles in the simulations. Here we fit directly to the baryonic masses so we remove the need to factor in the mass-to-light ratio degeneracies. We show the MGE fits in Fig. \ref{fig:allgal_contours.pdf} and see that for most galaxies the software outputs 2D gaussians that describe the baryonic distribution well. We note that in a couple of the galaxies the MGE's appear to be slightly off-centre, this is due to a wider asymmetry in the baryon distribution, but does not significantly affect the recovery of the enclosed mass profile. Secondly, the DM component is characterised by a generalised Navarro-Frenk-White (gNFW, \citet{Navarro1996}) profile \begin{equation}
\rho (r)=\frac{\rho_{0}}{({r}/{r_s})^{\gamma}(1+{r}/{r_{s}})^{3-\gamma}}, 
\end{equation}where $r$ represents the galactocentric spherical radius) where we leave the scale density ($\rho _ 0$), scale length ($r_s$) and inner slope ($\gamma$) as free parameters. This allows the profile to choose between cusps ($\gamma > 1$) and cores ($\gamma = 0$) while still becoming gNFW-like ($\rho \propto r^{-3}$) at large radii.
\item The measured line-of-sight velocities of each GC. Having the luxury of simulated data means we know the exact present day velocity for every GC and choose to use a 10$\kms$ line-of-sight velocity uncertainty. To obtain the GC line-of-sight velocity including observational errors we randomly sample a normal distribution with a mean of 0 km/s and a standard deviation of 10 km/s and add this to the true velocity from the simulation. We then include the 10 km/s uncertainty in the calculation of the likelihood. However, we discuss this choice and how different errors may affect the results in Sections \ref{6.2} and \ref{6.3}.
\item The distance and inclination angle of the galaxy. For this work we project our simulated galaxies at a distance of 1Mpc and at an inclination angle of 90 $^{\circ}$ (edge-on).
\item The rotation parameter ($\kappa$) of the GC system. This sets the relative contributions of random and ordered motion to the root mean squared velocities. In these models we assume the rotation parameter to be 0.
\item The velocity anisotropy parameter , \begin{equation}
\beta_{z} = 1 - \frac{\overline{v_{z}^{2}}}{\overline{v_{R}^{2}}},
\end{equation}calculated in cylindrical polar coordinates. We leave the velocity anisotropy to be free but constant i.e. a single value for the entire GC population. Parameterised this way, $\beta_z$  takes values in the range $- \infty$ (tangentially biased) to $1$ (radially biased). This large range of possible values can be difficult to work with. To deal with this, we re-parameterise as $\beta ' = {\beta}/({2-\beta})$, which takes values between -1 (vertical bias) and 1 (radial bias). 
\end{itemize}

We highlight here that we are using JAM models which assume cylindrical alignment of the velocity ellipsoid. An alternative would have been to consider models with a spherically aligned velocity ellipsoid. \citet{Cappellari2020} provides such JAM models along with accompanying software implementation. When choosing between the assumptions of cylindrical or spherical alignment, however, it is not a-priori obvious which choice is more suitable for our problem, despite our GC systems being close to spherical in their spatial distribution. This is because spherical distributions can still have cylindrically aligned velocity ellipsoids. Thankfully, there is reason to believe that this choice will not affect our results greatly. \citet{Cappellari2020} also shows that inferred density slopes are statistically indistinguishable when using the axisymmetric or spherically aligned assumptions. Since the main goal of this paper is to reconstruct mass profiles, this suggests that we would find similar results if we were to use the spherically aligned assumption.\par

To summarise, we have four free parameters in our model. These are three parameters from the DM distribution: the scale density,  the scale length and the inner slope ($\rho_0 , r_s, \gamma$) and a reparameterised version of the velocity anisotropy ($\beta '$). Because we have aligned our galaxies edge-on we remove the need for an inclination angle in the free parameters, but also adds an extra assumption to the rest of this work that should be kept in mind. However, for realistic disc galaxies, this work is best carried out in edge-on systems because these are the systems where the GC population is most readily observed and there are no uncertainties when deprojecting the MGE. We use a Markov-chain-Monte-Carlo (MCMC) method to explore this multi-dimensional parameter space.\par

\subsection{Markov-chain-Monte-Carlo\label{3.3}}
The major improvement to previous JAM models, made by \citet{Watkins2013}, is that there is no need to bin the data since the model takes discrete kinematic tracers. Previously, \citet{Gerssen2002} used individual stars with measured line-of-sight velocities to construct a discrete spherical Jeans model for the GC M15. \cite{Chaname2008} extended the Schwarzschild dynamic models to also include discrete data sets and \cite{vanderMarel2010} used line-of-sight velocities and proper motions of individual stars in the GC $\omega$-Centauri to find the presence of a possible intermediate-mass black hole in its centre. For small data sets such as the GC systems of MW-mass galaxies being able to use discrete data is a huge advantage because we can compare models against the data-set on a GC-by-GC basis by defining a discrete likelihood.
For this, we use Bayes' theorem, where the posterior probability distribution function of any free parameter in a model can be written as 
\begin{equation}
\mathrm{P}(\theta|D) = \frac{\mathrm{P}(D|\theta) \mathrm{P}(\theta)}{\mathrm{P}(D)},
\end{equation}
where $\theta=(\rho_{0},  r_s, \beta, \gamma)$ represents the model parameter set and $D=(x',y', v_{\mathrm{z'}})$ represents the data set. $\mathrm{P}(D|\theta)$ is the likelihood of an observation, given a model, $\mathrm{P}(\theta)$ is the prior and $\mathrm{P}(D)$ is a normalisation factor, sometimes called the evidence. The set of parameters that maximise the likelihood is the parameter set that reproduces the data most closely. In our case this means that this is the set of parameters that produce a DM halo that, together with the baryons, reproduces the GC velocities in the simulated galaxies.\par
The total likelihood of observing GC $i$ given model $\theta$ is the product of the model likelihoods for each individual GC:
\begin{equation}
\mathrm{P}(D|\theta) = \prod_{i=1}^{N_{\mathrm{GC}}} \frac{1}{\sqrt{2 \pi } \sigma_{i\theta}} \mathrm{exp} \left( -\frac{\left( \overline{v}_{\mathrm{z'},i} - \overline{v}_{\mathrm{z'},i\theta} \right) ^2}{2 \sigma_{i\theta}^{2}} \right),
\end{equation}
where
\begin{equation}
\sigma^2_{i\theta} = \overline{v^2}_{\mathrm{z'}, i\theta} - {\overline{v}_{\mathrm{z'},i\theta}}^2+ e^2,
\end{equation}
where $\overline{v}_{\mathrm{z'},i\theta}$ and $\overline{v^2}_{\mathrm{z'},i\theta}$ are the predicted first and second moments respectively, for a given GC position $i$, given the model parameters $\theta$. The error on the measurement of the true value is given by $e$ . We also note here that this equation is only rigorously correct if the line-of-sight velocity dispersion is described by a gaussian \citep{Mamon2013}.\par

We set flat priors given by:
\begin{itemize}
\item $5 \times 10^5   < \rho_{0} / [\Msun \kpc^{-1}] < 5 \times 10^7 $
\item $0   < R_{s}/[\kpc] < 50$
\item $0 < \gamma < 3$
\item $-1 < \beta ' < 1$
\end{itemize}

We explore parameter space using the Markov-Chain-Monte-Carlo (MCMC) package, EMCEE, developed by \citet{Foreman-Mackey2013}. 
EMCEE uses a number of independent walkers to explore the parameter space. Every walker takes a specified number of steps. At each step, an gNFW profile is calculated, for which a 1D spherical MGE can be fitted. This MGE is combined with the baryonic MGE (obtained by fitting the projected mass map directly from the simulation) to create the JAM potential density input. The JAM model then calculates the predicted first- and second-moments for the line-of-sight velocity at each GC position in this potential. We compare the JAM velocity moments with the true velocity moments (including the given velocity error) to return a likelihood value. We use 100 walkers, each taking 500 steps. MCMC converges well in 500 steps, we checked this by eye and saw that the walkers settled down to a stable state. We burn in the chain at step 50, therefore we use 450 steps to produce the posterior distributions.
\par

\section{JAM outputs \label{4}}
\subsection{Recovery of free parameters\label{4.1}}
\begin{figure*}
	\includegraphics[width=\linewidth]{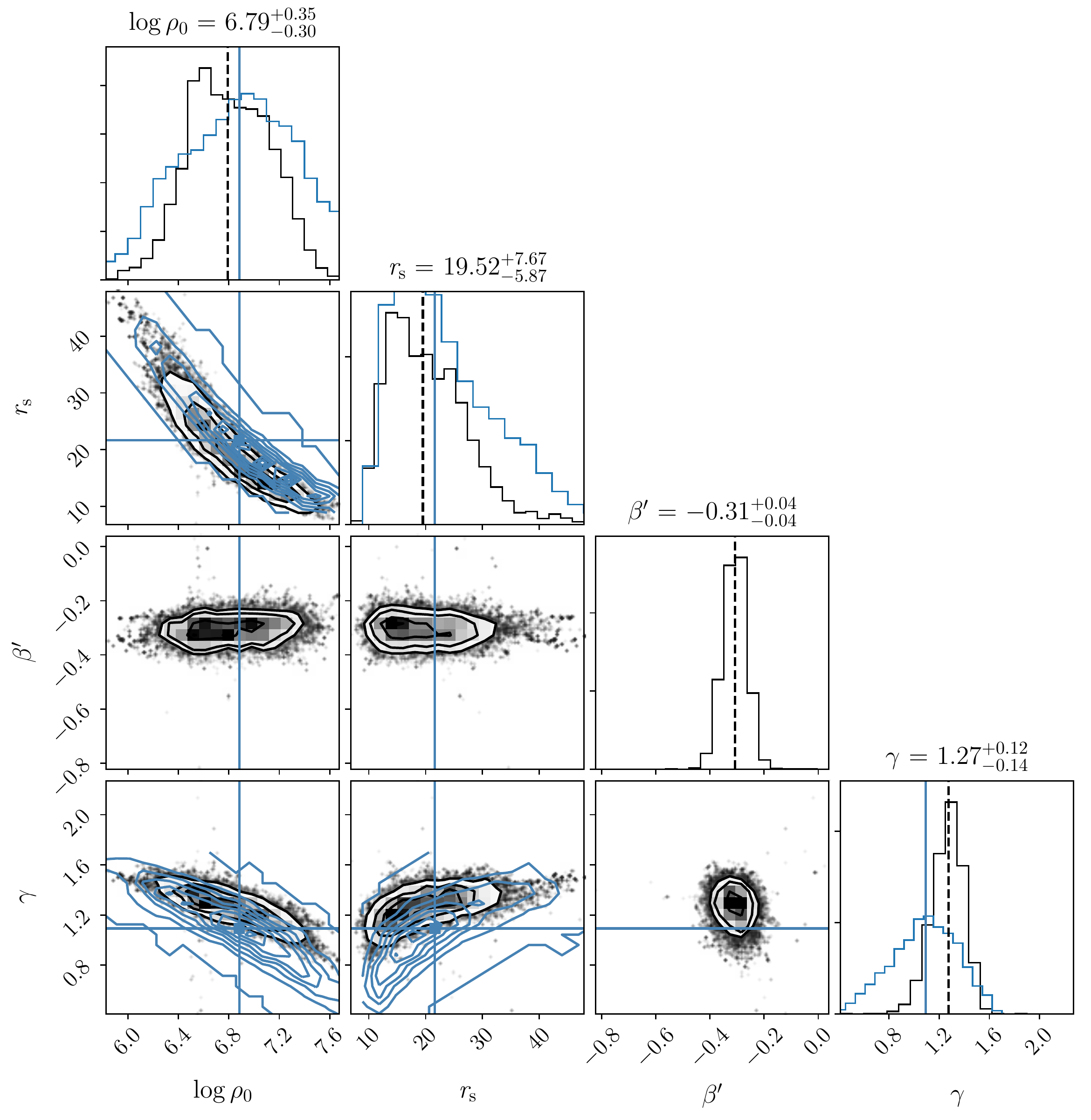}
    \caption{The posterior distributions for the four free parameters shown for one of the simulated galaxies (MW02). The grey 1D and 2D histograms represent the recovered values from the JAM model and the blue 1D histograms and vertical lines represent the fit directly to the simulated data.}
    \label{fig:Gal002_corner_2kpc.pdf}
\end{figure*}

\begin{figure*}
	\includegraphics[width=\linewidth]{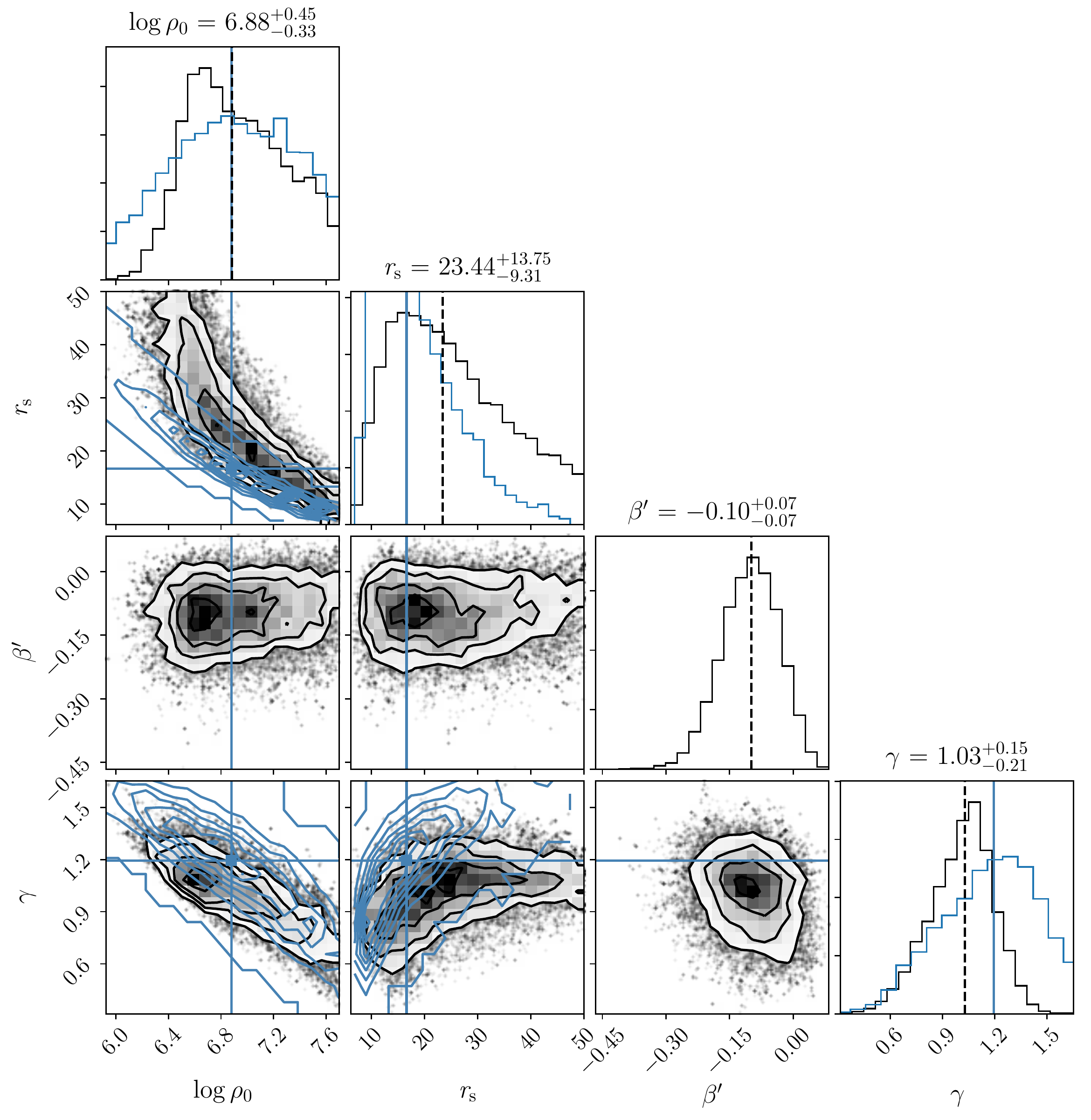}
    \caption{The posterior distributions for the four free parameters shown for one of the simulated galaxies (MW04). The grey 1D and 2D histograms represent the recovered values from the JAM model and the blue 1D histograms and vertical lines represent the fit directly to the simulated data.}
    \label{fig:Gal004_corner_2kpc.pdf}
\end{figure*}

\begin{figure}
	\includegraphics[width=\linewidth]{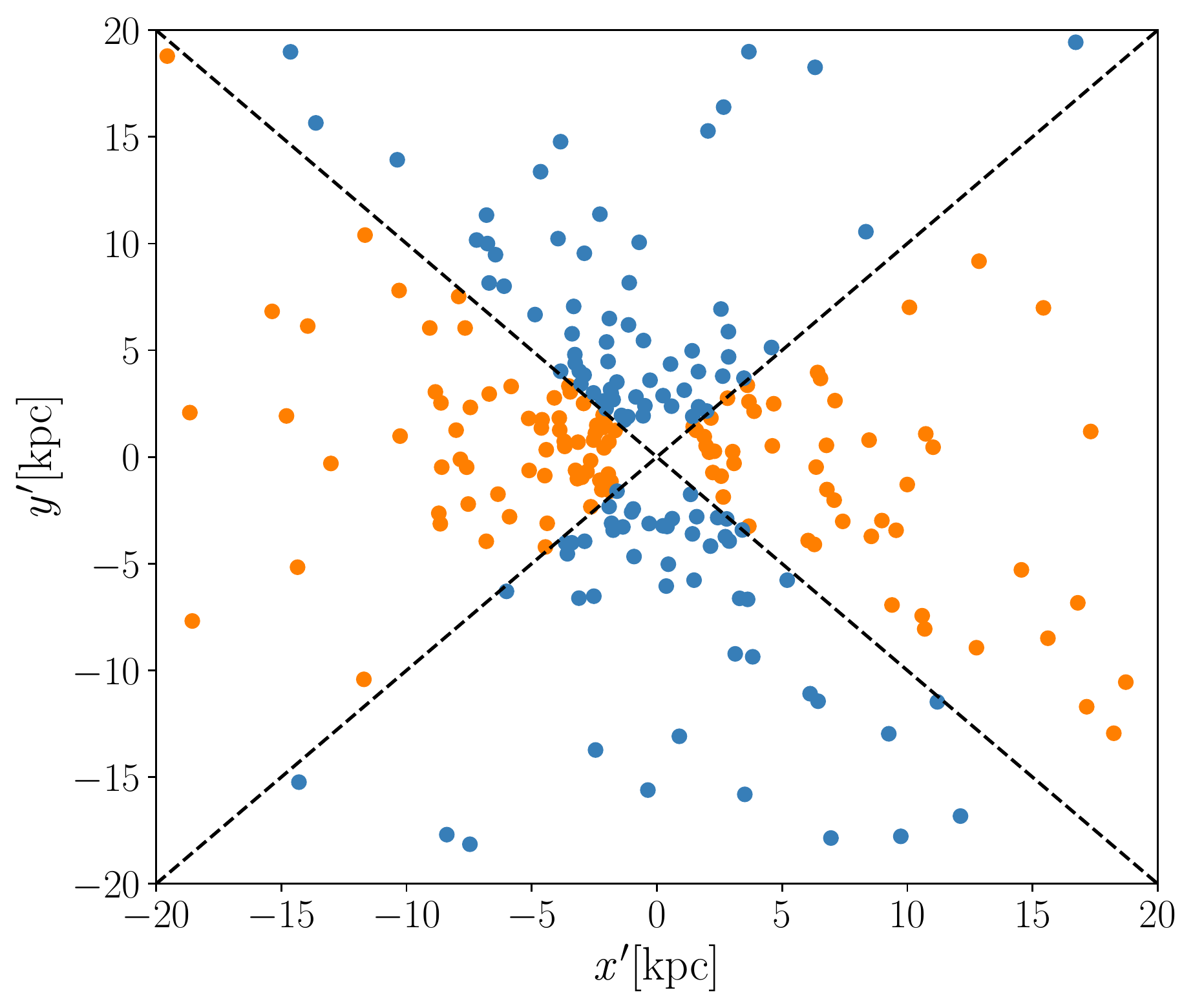}
    \caption{A demonstration of how we assign GCs to the major or minor axes for MW16. Each point represents one GC, coloured by whether it belongs to the major or minor axis population, with the minor axis shown in blue and the major axis shown in orange. The void in the centre is due to our inner 2 kpc radius cut.}
    \label{fig:GCpos.pdf}
\end{figure}

\begin{figure*}
	\includegraphics[width=\linewidth]{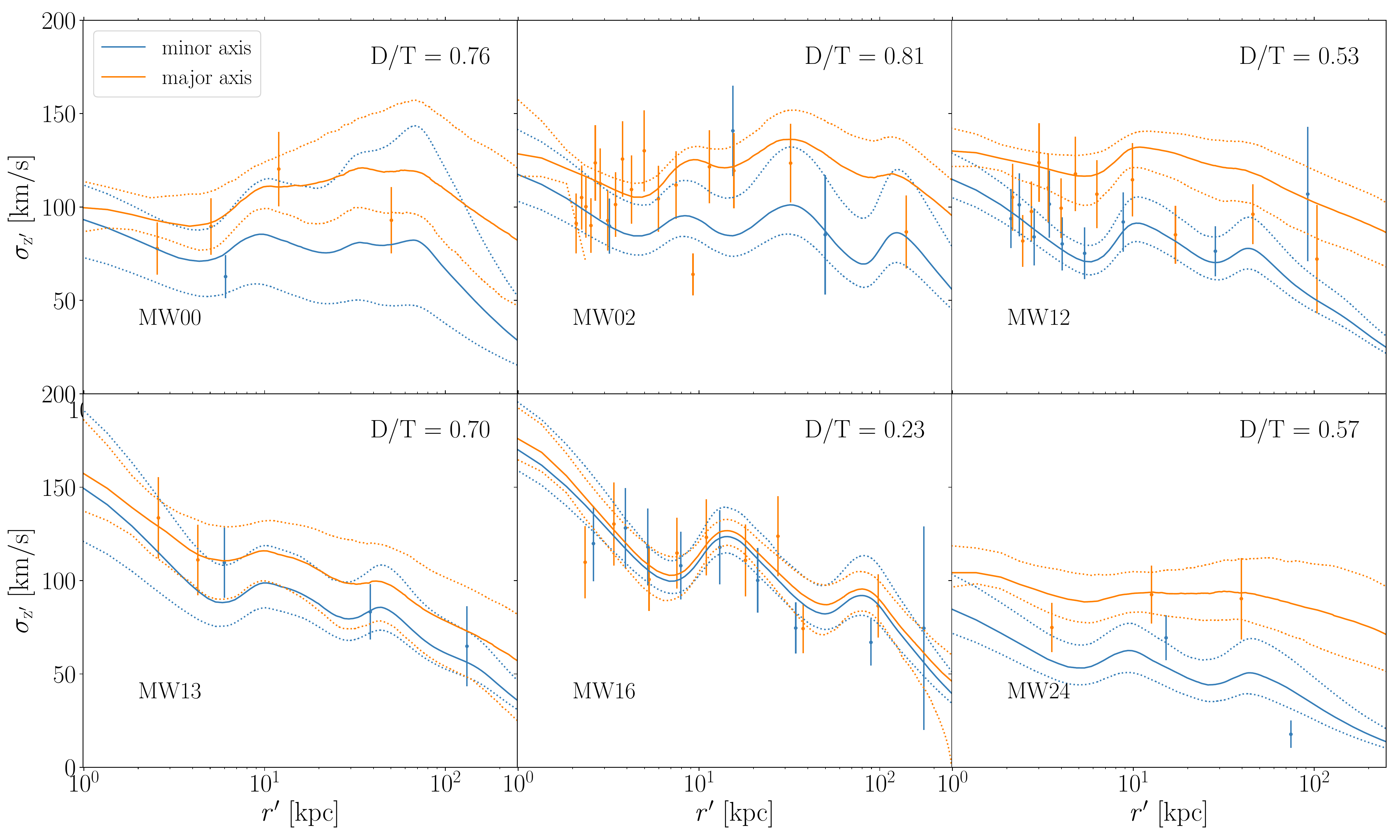}
    \caption{The line of sight velocity dispersion ($\sigma_{\mathrm{z'}}$) recovered from JAM and directly from the simulations as a function of projected radius for 6 of the simulated galaxies. The continuous solid line shows the median JAM output and the dotted lines represent the 1$\sigma$ spread in the predictions. The points with error bars show the true $\sigma_{\mathrm{z'}}$ calculated directly from the simulations. To calculate the true GC $\sigma_{\mathrm{z'}}$ we binned the GCs in groups of 20. In the top right corner we indicate the disc to total mass ratio (D/T) for each galaxy.}
    \label{fig:veltest_minormajor.pdf}
\end{figure*}

As discussed in Section \ref{3.3} the JAM model, when combined with an MCMC, returns a posterior distribution of the four free parameters in the model. These are three free parameters in the gNFW halo and the velocity anisotropy of the GCs. To understand whether the JAM model is performing well we would like to compare our results to the 'true' values of the free parameters in the gNFW halo. However, we must also acknowledge that a range of parameter combinations may produce the same DM radial profile. To quantify this degeneracy in the `true' parameter values, we perform an MCMC parameter exploration directly on density profiles from the simulation.

We first calculate a 1D DM density profile directly from the DM particles in the simulation using 100 logarithmically-spaced spherical shells. We then fit a gNFW profile  to the true density distribution using the same free parameters, priors, number of walkers and number of steps as used for the JAM model. This then gives us a posterior distribution for the true parameters that we can directly compare to the posterior distribution from the model. We include this step because the E-MOSAICS simulations are fully cosmological simulations and therefore their density profiles are only approximated by the gNFW parameterisation, so there will always be some uncertainty in the gNFW fit.\par

The 1D and 2D posterior distributions for two galaxies (MW02 and MW04) are shown in Figs. \ref{fig:Gal002_corner_2kpc.pdf} and \ref{fig:Gal004_corner_2kpc.pdf} where the black data shows the outputs from the JAM model and the blue data shows the outputs from the gNFW fit applied directly to the DM particles from the simulation. We exclude a small percentage of the walkers from this analysis as they diverged from the majority. The first thing we note is that there are significant degeneracies in the recovery of the true parameters: $\log{\rho_{\mathrm{0}}}$ anti-correlates with $r_\mathrm{s}$ and $\gamma$, and $r_{\mathrm{s}}$ correlates with $\gamma$. This means that although there are most-likely values, there are multiple combinations of these parameters that will produce the same fit to the DM particles. Therefore, we should be wary of comparing the absolute values of the free parameters given by the JAM model and the fit directly to the DM particles. We discuss our chosen method of comparison in the next section.\par

In the contour plot for MW02 (Fig. \ref{fig:Gal002_corner_2kpc.pdf}) we see that the posterior distributions for the JAM model almost lie on top of that for the true fit, meaning that the recovered values for each DM halo parameter are close to that of the true value and that the degeneracies encountered in the JAM model are explained by the intrinsic degeneracies in the parameters. The 1D histograms show the posterior distributions for each parameter individually. The blue histogram and solid blue vertical line represent the posterior distribution and the median value of the `true' fit and the black histogram and vertical dotted line represents the same for the JAM fit. In Fig. \ref{fig:Gal002_corner_2kpc.pdf} the posterior distributions for the true fits and the JAM model are similar in shape and the median values are reasonably close. However, for MW04 (Fig. \ref{fig:Gal004_corner_2kpc.pdf}) the posteriors for the JAM model are offset slightly from those of the true fit, this is also evident in the shape of the posterior distributions for each parameter, although the median values for the free parameters are still recovered well. We also note that the posterior distributions are wider in shape for MW04 (Fig. \ref{fig:Gal004_corner_2kpc.pdf}) than for MW02 (Fig. \ref{fig:Gal002_corner_2kpc.pdf}), which means that the JAM model cannot get as good constraints on the DM distribution in MW04. These two galaxies were chosen because they represent two different cases of the DM distribution recovery.  Table \ref{table:GCnumber} shows that MW04 has almost 4 times fewer GCs than MW02 and therefore this could be the driving factor for the difference in the constraints for the two galaxies. This is discussed further in Section \ref{6.1}, but for now this is a good demonstration of how well the JAM model is performing in general and, after inspecting all the recovered parameters for all galaxies, there are no JAM recoveries that differ too greatly from the true recovered parameters.\par

\subsection{Recovery of velocity moments\label{4.2}}
The JAM model calculates the first and second velocity moments at any given position in the galaxy. We can therefore compare the line of sight velocity dispersion ($\sigma_{\mathrm{z'}}$) as given by JAM to that calculated directly from the simulations. Since this is an axisymmetric model, we first divide the galaxy into a major ($x'$) and minor ($y'$) axis. For the JAM predictions, this is relatively straightforward since we just calculate the velocity moments directly along the $y' = 0$ axis and the $x' = 0$ axis to obtain predictions for the minor and major axis respectively. We divide the x-y plane into quadrants using the $x=y$ and $x=-y$ lines, and assign GCs to the major or minor axis according to the quadrant where they are located, as demonstrated by  Fig. \ref{fig:GCpos.pdf}. The blue points represent the minor axis GCs and the orange points represent the major axis GCs. The void in the centre is our 2$\kpc$ inner radius cut.\par

The calculation of the true first- and second-moments of the velocity directly from the simulations is reasonably straightforward. The first moment is simply the $v_{z'}$ outputted directly from the simulation. However, for the second moment we must bin the GCs. The GCs are binned along each axis in groups of 20 and the second moment is given by
\begin{equation}
\overline{v^{2}}_{z'}=\frac{1}{N_{\mathrm{GC}}} \sum^{N_{\mathrm{GC}}}_{i=1} \overline{v}_{z',i}^{2}
\end{equation}
where $N_{\mathrm{GC}}$ is the number of GCs in the bin. From the first- and second-moments the final line-of-sight velocity dispersion is given by 
\begin{equation}
\sigma_{\mathrm{z'}} = \sqrt{\overline{v^{2}}_{z'} - \overline{v}_{z'}^{2}}.
\end{equation}
The JAM predictions and simulation calculations for $\sigma_{\mathrm{z'}}$ are shown for six galaxies in Fig. \ref{fig:veltest_minormajor.pdf}, where the solid and dotted lines represent the median and 1$\sigma$ spreads from the JAM model respectively and the points represent the velocity dispersion calculated directly from the simulation. Each point is calculated using 20 GCs, with the x-axis position representing the centre of the 20 GC bin (for example we see that MW02 has many GCs in its central 10$\kpc$). We fold the galaxy about the minor axis, this is done because the first moment of the velocity is squared, so it removes the need to keep the sign of the velocity. The error bars are calculated via a Monte-Carlo error calculation.\par

Fig. \ref{fig:veltest_minormajor.pdf} shows that the $\sigma_{\mathrm{z'}}$ from the JAM model matches that calculated from the simulations reasonably well within the errors. This is a good demonstration of how we can construct good dynamical models of galaxies using just the GCs as tracers, even for galaxies such as MW13 where there are a limited number of tracers. We do notice however that in MW12, although the JAM model does a good job at predicting the velocity dispersion along the minor axis it over predicts the velocity dispersion along the major axis particularly in the outskirts. \par

Fig. \ref{fig:veltest_minormajor.pdf} also shows that there is variation in the shape of the JAM predictions between galaxies. We particularly take note of the difference between some galaxies, such as MW16 and MW12, where the predictions for the minor and major axis are very similar and reasonably different, respectively. The number in the top right corner of each panel represents the disc to total mass ratio (D/T). D/T is calculated by assigning stellar particles to the disc or halo component of the galaxy. This is done by calculating the fraction of angular momentum that is in the disc plane for each particle, also known as the circularity parameter $\epsilon_\mathrm{J} = J_\mathrm{z}/J_\mathrm{c} (E)$  \citep{Abadi2003} where $\epsilon_{\mathrm{J}} = 1$ describes a perfectly circular orbit. The stellar particles that belong to the disc component have $\epsilon_{\mathrm{J}}>0.5$ \citep{Sales2015}.

MW16 is likely to have velocity dispersion profiles that are similar along the minor and major axis due to the fact that this galaxy is not very disc dominated and is therefore more elliptical in shape, making the galaxy more spherically symmetric. In addition, MW16 also has a similar number of tracers along its major and minor axis so the data quality is similar in both cases. The rest of the galaxies in Fig. \ref{fig:veltest_minormajor.pdf} have D/T > 0.5 and therefore are in the disc-dominated regime. In these cases there is a clear difference between the major and the minor axis velocity dispersions, with the major axis being dominated by the disc component always having higher velocity dispersion. This is because, in a disc galaxy orientated edge-on the major axis constitutes a thicker component and there are particles on different stages of their orbits.\par

%\textcolor{red}{relate this back to the velocity anisotropy. For an axisymmetric system following the assumptions of the JAM models, the information of velocity anisotropy is encoded in the difference of velocity dispersion along the major and minor axes. How is the velocity anisotropy predictions different for these galaxies?}

\subsection{Recovery of dark matter mass distribution\label{4.3}}

\begin{figure*}
	\includegraphics[width=\linewidth]{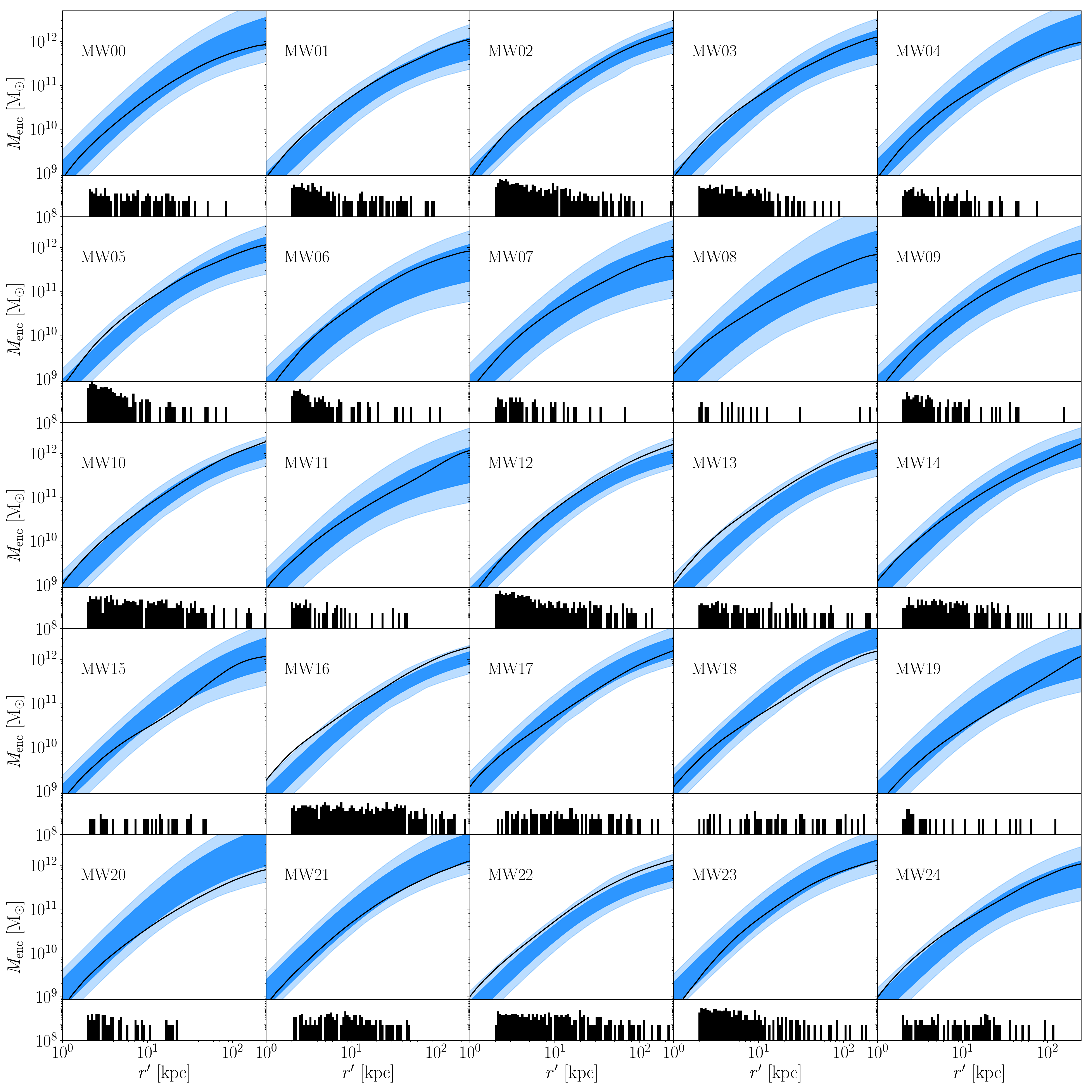}
    \caption{Projected radial profiles of the mass enclosed for the 25 galaxies, using all clusters at $z=0$ with a mass $> 10^5 \mathrm{M_{\odot}}$, an age $> 8 \mathrm{Gyr}$ and a galactocentric radius R>2 kpc. The solid black line represents the true mass calculated directly from the simulations. The blue-shaded regions represent the $1$ and $2 \sigma$ bounds on the recovered mass from the JAM model. Each panel also contains a histogram of the 2D projected GC positions.}
    \label{fig:mass_agecut_2kpc.pdf}
\end{figure*}

\begin{figure}
	\includegraphics[width=\linewidth]{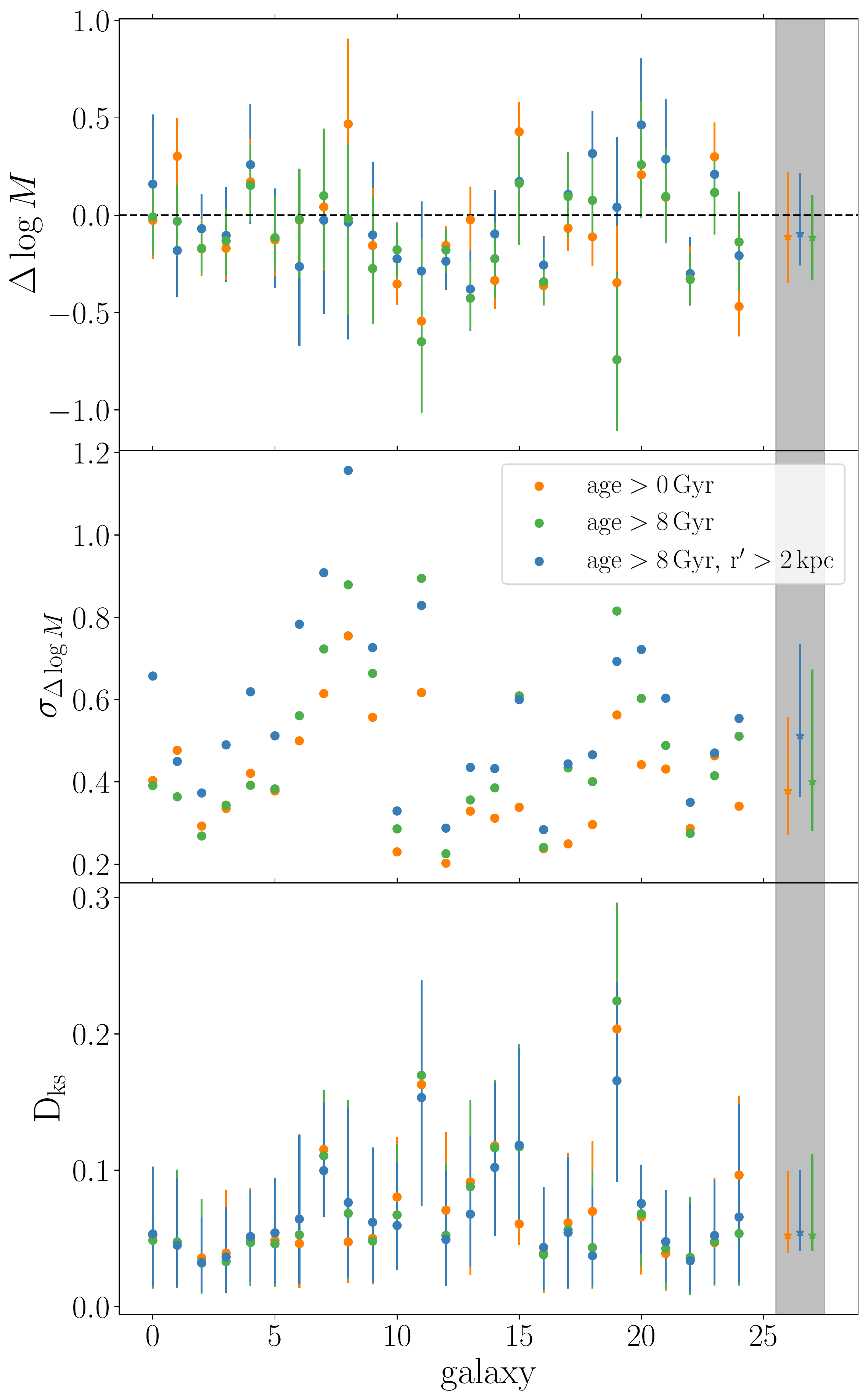}
    \caption{Estimators of the performance of the JAM model. From top to bottom, they quantify the DM mass difference, the spread in the DM mass difference and the maximum values of the KS test, for all the galaxies in the suite of 25 Milky Way-mass zoom simulations. Data points correspond to different samples of GC tracers as indicated in the legend. The grey-shaded region contains the averages and $1\sigma$ spreads for each of the GC sub-samples, indicated by the stars with errorbars.}
    \label{fig:ks_mass_agetest_newmass.pdf}
\end{figure}

As it is shown and discussed in Section 4.1, there are significant degeneracies between the free parameters in the gNFW profile. Therefore it is better not to compare the true and JAM-recovered individual parameters, but to compare the DM distribution and mass enclosed within a chosen radius given by the simulations and recovered by the JAM model. This mitigates the need to take parameter degeneracies into account.  We therefore convert the gNFW profile parameters into enclosed mass profiles. For the JAM outputs we calculate a realisation of the profile for each MCMC run and plot the $1$ and $2 \sigma$ spread of the mass enclosed at each radius. The DM enclosed mass profiles for the 25 simulated MW-mass galaxies are shown in Fig. \ref{fig:mass_agecut_2kpc.pdf}, where the black solid line represents the DM mass calculated directly from the DM particles in the simulation. The blue shaded regions represent the $1$ and $2 \sigma$ recovered DM mass from the JAM model. At the bottom of each panel, we also show a histogram of the GC positions, plotted as a function of projected spherical coordinate $R'$. This is useful to understand how the GCs are distributed in 2D projected $x'-y'$ space. Fig. \ref{fig:mass_agecut_2kpc.pdf} shows that the radial distributions of the recovered and true DM distributions are similar for all galaxies, with no obvious outliers. However, we do notice that there are clear differences between how well the JAM model performs for each galaxy. For example, when comparing MW16 with MW07, the JAM model under predicts the DM mass in MW16 but the $1$ and $2 \sigma$ spread are very small. However, the true DM mass in MW07 is almost in the middle of the $1 \sigma$ spread, its precision is much larger. To compare quantitatively the results for different simulated galaxies we use three quantities: the recovered vs the true mass enclosed at the maximum GC radius, the $1 \sigma$ spread in this mass and a cumulative distribution test.

To quantify the difference between the recovered DM mass from JAM and the true DM mass within the maximum GC radius we use the log difference, 
\begin{equation}
\Delta \log{M}=\mathrm{log}(M_{\mathrm{JAM}}/M_{\mathrm{true}}),
\end{equation}
where $M_\mathrm{JAM}$ is the median of the posterior of the JAM fits. This means that JAM model over- or under-predicts the mass when $\logm > 0 $ and $\logm< 0$, respectively. 

The second test we use is the spread in $\logm$ due to the $1 \sigma$ spread in the recovered mass from JAM, 
\begin{equation}
\sigma_{\Delta \log{M}} = \mathrm{log}(M_{\mathrm{JAM, 84}}/M_{\mathrm{true}}) - \mathrm{log}(M_{\mathrm{JAM, 16}}/M_{\mathrm{true}}) 
\end{equation}
where $M_{\mathrm{JAM,16}}$ and $M_{\mathrm{JAM,84}}$ represent the 16th and 84th percentiles of the JAM-recovered enclosed mass. These two values inform us about how close to the true value and how well constrained the JAM-recovered mass is. \par

The previous two parameters inform us about the mass of DM at a given radius. However, the free parameters in the JAM model also describe the radial profile of the DM distribution, therefore we also test how the radial profile of the DM from the JAM model compares to the true profile. For this, we use a Kolmogorov-Smirnov (KS) test, which is a nonparametric test of whether two cumulative distributions differ. The KS test determines whether the null-hypothesis that the two distributions are from independent samples drawn from the same underlying form is likely to be true or not. The KS D value ($\mathrm{D_{KS}}$) represents the maximum value of the absolute difference between the two cumulative distribution functions and therefore the lower this number, the more likely it is that the two distributions match. We calculate $\mathrm{D_{KS}}$ for each simulated galaxy by calculating the difference between the normalised cumulative distribution functions of the DM radial profiles directly from the simulations and the JAM output. We do this at 100 radii logarithmically spaced between 2 kpc and the maximum GC radius, and find the maximum difference.\par

To summarise, we have now described three parameters that we use to quantify how well the JAM model performs, two that describe how well the mass at the maximum GC radius is recovered ($\logm$, $\sigma_{\logm}$) and one that quantifies how well we recover the radial profile of the DM distribution ($\mathrm{D_{KS}}$).\par

Fig. \ref{fig:ks_mass_agetest_newmass.pdf} shows these three parameters for all 25 galaxies. We show, for each galaxy, three variations on the selection of GCs (which we include in all steps of the model setup, and given in Table \ref{table:GCnumber}). The blue points represent the case where we use all of the star clusters with $M>10^5 \Msun$, the orange points represent the case where we restrict the sample to the classical ancient GCs ($M>10^5 \Msun$, age $> 8 \Gyr$) and the green points represent the case where we also exclude the inner-most GCs ($M>10^5 \Msun$, age $> 8 \Gyr$, $r' > 2 \kpc$). The motivation for these cuts is described in Section \ref{2}. In Fig. \ref{fig:ks_mass_agetest_newmass.pdf} we also show the median of each of the three values, represented by the stars within the grey shaded region. We see that in all GC sub-samples the JAM model marginally underestimates the DM mass, but is consistent with not showing any biases.
In the middle panel of Fig. \ref{fig:ks_mass_agetest_newmass.pdf} we see that the subset of GCs with $M>10^5 \Msun$, age $> 8 \Gyr$, and $r' > 2 \kpc$ have the largest $\sigma_{\Delta \log{M}} $. However, in the $D_{\mathrm{KS}}$ test, which quantifies the recovery of the DM radial distribution, none of the three GC sub-samples performs consistently better than the others. The radial profile recovery is reasonably consistent between galaxies, with two potential outliers. Galaxies MW11 and MW19 (along with MW15) have a massive companion galaxy that is distorting the shape of the DM halo and causing it to no longer be well paramaterised by an gNFW halo. The mass of the DM halo is still well recovered in these galaxies and we therefore keep them in the rest of the analysis.\par

For the rest of the analysis we use the GC selection that includes the radius and the age cut, motivated by previous discussions about observational constraints. Therefore, in the rest of this work, we define a GC as a star cluster with $M > 10^{5} \Msun $, age$>8 \Gyr$ and $r'> 2 \kpc$. We now look into the different properties of the GC populations and investigate which galaxy properties may be affecting the variations in the recovery of the DM profile.

\section{Investigation of the effects of the GC system}\label{5}
\subsection{Effect of GC distribution sphericity}
\begin{figure}
	\includegraphics[width=\linewidth]{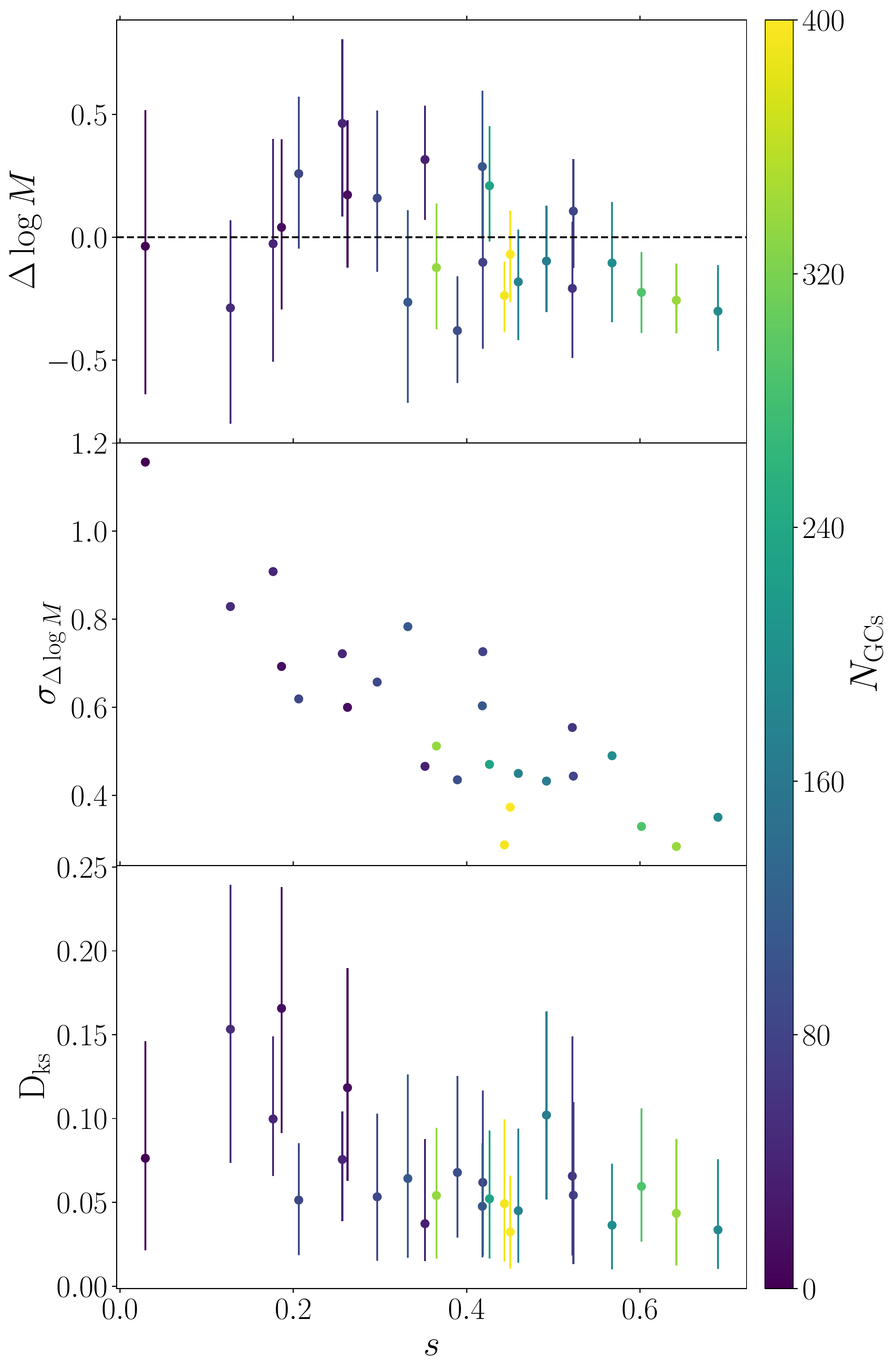}
    \caption{Estimators of the performance of the JAM model as in Fig. \ref{fig:ks_mass_agetest_newmass.pdf}, as a function of the sphericity ($s$) of the GC distribution, with $s=1$ being a spherical distribution. The points are also coloured by the number of GCs used in the JAM model.}
    \label{fig:ks_mass_sphericity.pdf}
\end{figure}
In the JAM model we force the GC distribution to be spherical because we model it with a spherical MGE. However, it is interesting to see how spherical these distributions are in the simulations and how deviations from sphericity affects the recovery of the DM distribution.\par
We obtain a `sphericity' parameter ($s$) for the GC spatial distribution using the methodology described in \citet{Thob2019}. The 3D spatial distribution of GCs is modelled with an ellipsoid with major, intermediate and minor axes $a$, $b$ and $c$. The sphericity of this distribution can then be defined as,
\begin{equation}
s=\frac{c}{a}.
\end{equation}
For spherical distributions, $s=1$. The axis lengths are defined by the square root of the eigenvalues of a matrix that describes the three-dimensional GC distribution. The matrix is chosen to be the tensor of the quadrupole moments of the spatial distribution (often referred to as the moment of inertia tensor). We direct the interested reader to \citet{Thob2019} for more details.\par

In Fig. \ref{fig:ks_mass_sphericity.pdf} we show the mass and radial profile recovery of the DM distribution as a function of $s$. Each point represents one galaxy and the points are coloured by the number of GCs in each galaxy. First of all, we note that although we are modelling the GC distributions as spherical, they often deviate from this assumption. This could be due to sparse sampling in the calculation of this parameter. We perform Monte Carlo simulations to estimate the axis ratio of 2D ellipsoids and find that the error will increase with increasing sphericity. We calculate a maximum error of 0.06 for s>0.6. \par

We first focus on the recovered mass ($\logm$ and $\sigma_{\logm}$). Importantly, the sphericity does not have an effect on the absolute value of $\logm$ i.e. the sphericity does not bias the recovery of the mass enclosed. As expected, as $s$ increases (the GC distribution becomes more spherical), the recovered mass becomes more constrained, as shown by $\sigma_{\logm}$ becoming smaller.  Next, we focus on the radial profile of the distributions, $\mathrm{D_{KS}}$ shows that as $s$ tends towards 1 the radial profile of the recovered DM becomes closer to the true radial profile. However, those galaxies with GC systems with a sphericity parameter greater than $\approx 0.3$ have similar $\mathrm{D_{KS}}$ values. \par

Fig. \ref{fig:ks_mass_sphericity.pdf} shows us that there tends to be more GCs in more spherical systems. For example, the galaxy that has the least spherical GC distribution (MW18) also has the least number of GCs, at just 18. This is potentially related to another correlation between assembly history and richness of the GC system, with galaxies with fewer minor mergers have fewer GCs \citep{Kruijssen2019}. Therefore, the galaxies with fewer GCs also underwent fewer minor mergers and therefore could not uniformly populate the halo volume with GCs. We would expect that galaxies with fewer GCs would also have flatter distributions because their GC population is made up of a higher fraction of in-situ GCs which formed in the disc. Above a sphericity of 0.3 it seems that the two properties of the GC system (number and sphericity) work together to affect the JAM model. $\sigma_{\logm}$ is, on average, continuously decreasing with $s$, however at a given $s$ (above 0.3) $\sigma_{\logm}$ depends on the number of GCs. This is also apparent in the radial profile recovery -- the two galaxies that have the best match in the DM distribution are also the two galaxies with the most GCs.\par

Therefore, we conclude that both the sphericity and the number of GCs impact how well the DM profile can be recovered. However, within the models, we note a strong correlation between the sphericity and GC number, hence observations limited to large GC systems are less likely to be affected by non-sphericity of their GC systems. We investigate the impact of the number of GCs in the next section.\par

We also note here that we find that the concentration of the GC system ($r'_{\mathrm{median}}/r'_{\mathrm{max}}$) plays no role in the recovery of the DM mass or overall radial profile.

\subsection{Number of GCs\label{6.1}}
\begin{figure*}
	\includegraphics[width=\linewidth]{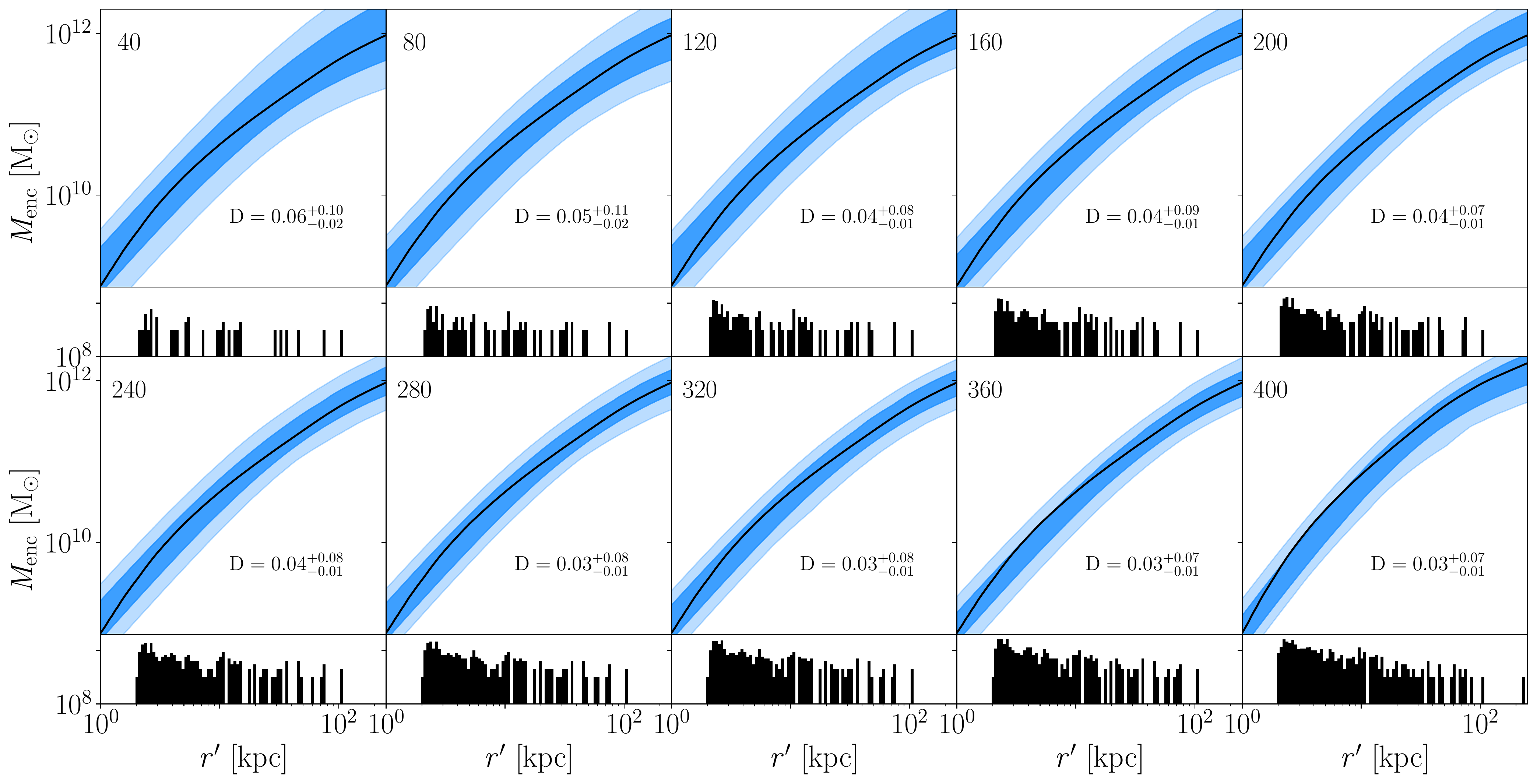}
    \caption{Mass enclosed profiles for MW002 with it's GCs randomly sampled to include 10-100 percent of them. We show the $\mathrm{D_{KS}}$ value of each JAM model in the relevant panel. The number in the top left of each panel shows the number of GCs used in the JAM model. The lines and shaded regions correspond to the description in Fig. \ref{fig:mass_agecut_2kpc.pdf}}
    \label{fig:Gal002_mass_numberTest.pdf}
\end{figure*}

\begin{figure}
	\includegraphics[width=\linewidth]{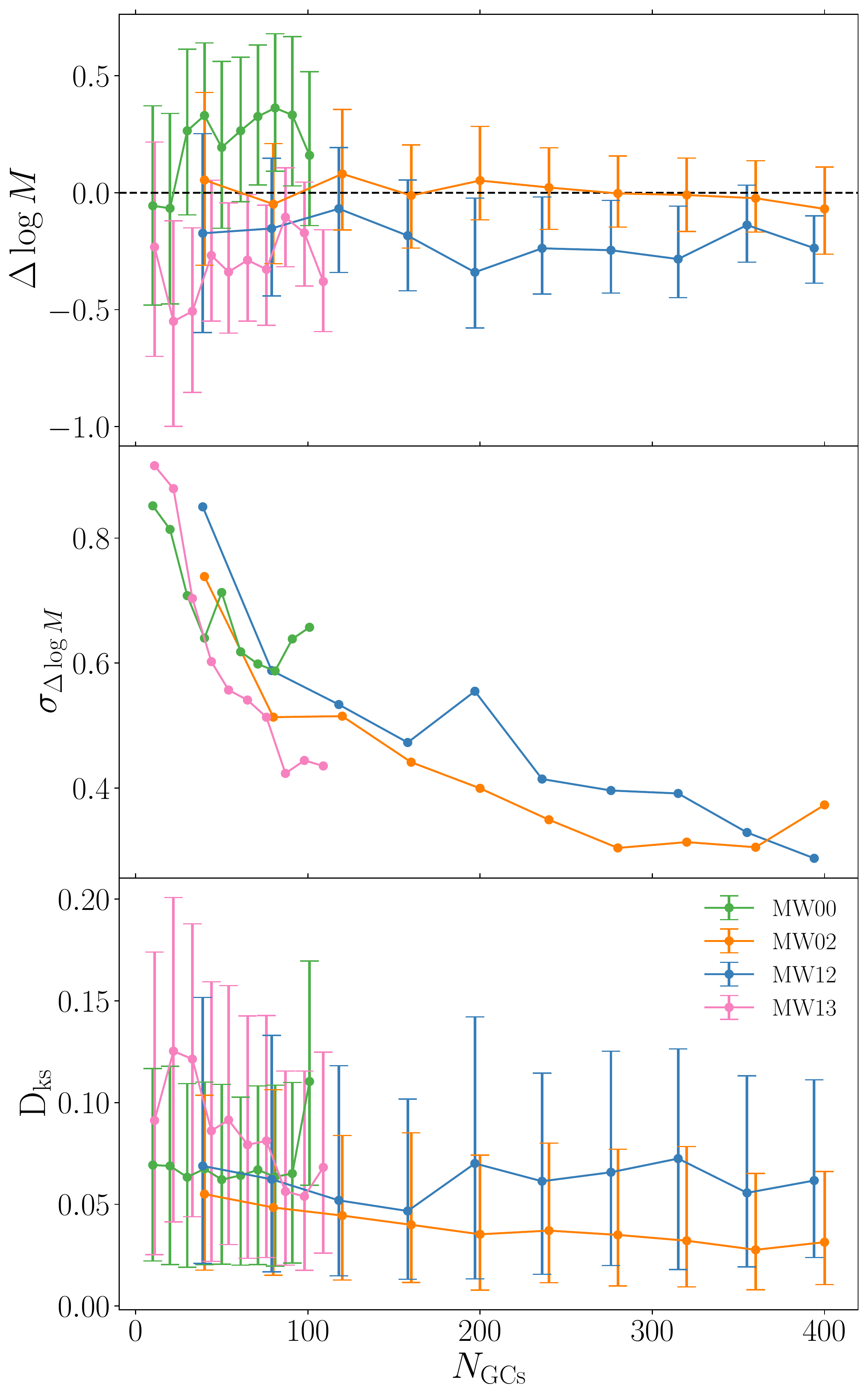}
    \caption{Estimators of the performance of the JAM model as in Fig. \ref{fig:ks_mass_agetest_newmass.pdf}, as a function of the number of GCs used in the model. Here we show four galaxies: MW00, MW02, MW12 and MW13 indicated by the legend.}
    \label{fig:ks_mass_numberTest_GCnumber.pdf}
\end{figure}

\begin{figure}
	\includegraphics[width=\linewidth]{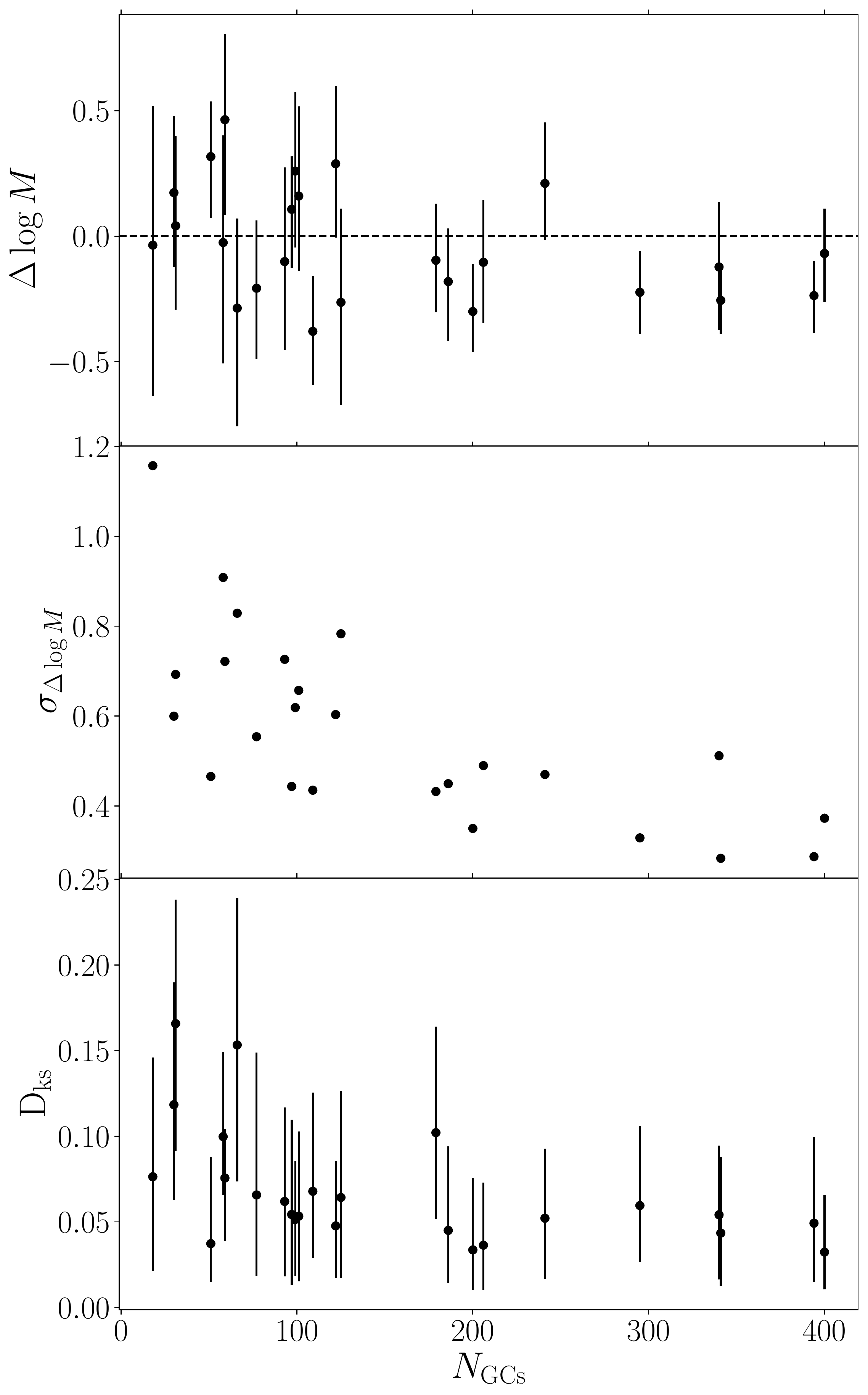}
    \caption{Estimators of the performance of the JAM model as in Fig. \ref{fig:ks_mass_agetest_newmass.pdf}, as a function of the total number of GCs in the galaxy and used in the JAM model. Each point represents one of the 25 simulated galaxies.}
    \label{fig:ks_mass_numGCs.pdf}
\end{figure}

 We now investigate how the number of dynamical tracers used as an input to the JAM model affects how well we recover the DM mass profile. We do this in two ways: first by varying the number of GCs in one galaxy and then a comparison between the number of GCs in different galaxies. Note that in this subsection we keep the velocity error $\Delta v_{z'} = 10\kms$.\par

First we randomly sample the GCs in one galaxy, to select 10, 20, 30, 40, 50, 60, 70, 80 and 90 per cent of their GCs as inputs to the JAM model. We use just the randomly sampled GC throughout the whole JAM model, including the initial MGE fit to their positions. We then carry out our analysis in the same way as before. Fig. \ref{fig:Gal002_mass_numberTest.pdf} shows the enclosed mass profiles for MW02 for each subset of GCs. Here the plots are the same as described for Fig. \ref{fig:mass_agecut_2kpc.pdf}. From left to right and top to bottom the number of GCs used is increased in steps of 10 per cent and the number in the top left corner is the number of GCs used. In the bottom right of each panel the KS statistic is quoted. As can be seen in the top panels, the uncertainty in the recovered DM radial profile decreases as the number of GCs is increased from 40 to 200. Not much further improvement is gained by increasing the number of GCs to 400 (bottom panels). It is particularly encouraging that in this galaxy, the radial profile is well recovered even with only 40 GCs, with errors less than 6 per cent.\par

We follow the same method of randomly sampling GCs in three more galaxies. We chose these simulated galaxies because their mass profiles are recovered similarly well when all their GCs are used, but they have a different total number of GCs. The mass and radial profile recovery for these four galaxies, as a function of the number of GCs used ($N_{\mathrm{GCs}}$) is shown in Fig. \ref{fig:ks_mass_numberTest_GCnumber.pdf}.  The recovery of the radial profile of the DM improves when a higher fraction of a galaxy's GCs is used, as shown by the average $\mathrm{D_{KS}}$ value decreasing with increasing $N_{\mathrm{GCs}}$. There is no systematic trend in $\logm$ with the fraction of GCs used, but there is a consistent offset for each galaxy, meaning that if the model underestimates the mass when using 100 per cent of its GCs, it does when using a smaller fraction of its GCs. This shows that the number of GCs does not cause an estimate of the enclosed DM mass to be an over- or under-estimation. The one sigma spread ($\sigma_{\logm}$) decreases as the fraction of GCs increases. This forms a narrow negative correlation. An increasing number of GCs are needed to get the JAM models as well constrained and as close to the shape of the true DM profile as possible. \par

We now plot every galaxy as a function of the total number of GCs used in the Jeans model. This is shown in Fig. \ref{fig:ks_mass_numGCs.pdf}, where each point now represents an individual galaxy, with all of its GCs. Here we see that the Jeans models produce consistently lower $\sigma_{\logm}$ in galaxies with more than 150 GCs compared to those with fewer than 150 GCs. The Jeans models in the galaxies with increased number of GCs also recover the overall radial profile of the DM consistently well. However, it does not mean that in galaxies with fewer than 150 GCs, models systematically fail at recovering the mass and radial distribution of the halo, but instead that there is a lot of scatter in the recovered mass and profile. To guarantee that the JAM model will perform well, we would suggest that more than 150 GCs are needed for a Jeans model of a Milky Way mass galaxy. As previously mentioned, the mean accuracy in the recovered mass ($\logm$) for all the galaxies is 0.21 dex with a precison ($\sigma_{\logm}$) of 0.57 dex. However, when considering galaxies with more than 150 GCs, although the accuracy only improves by 0.03 to 0.18 dex, the precision increases to 0.38 dex, almost a 0.2 dex improvement in precision. We conclude that with fewer than 150 GCs, the Jeans model could be recovering the mass profile accurately without bias. However, to obtain a precise constraint more than 150 GCs are needed.\par

The analysis so far assumed a line-of-sight velocity error of $10\kms$. However, the number of GCs needed may be impacted by the quality of the data. We therefore investigate the effect of data quality by increasing the line-of-sight velocity error in the next section.

\subsection{Line-of-sight velocity error\label{6.2}}

\begin{figure}
	\includegraphics[width=\linewidth]{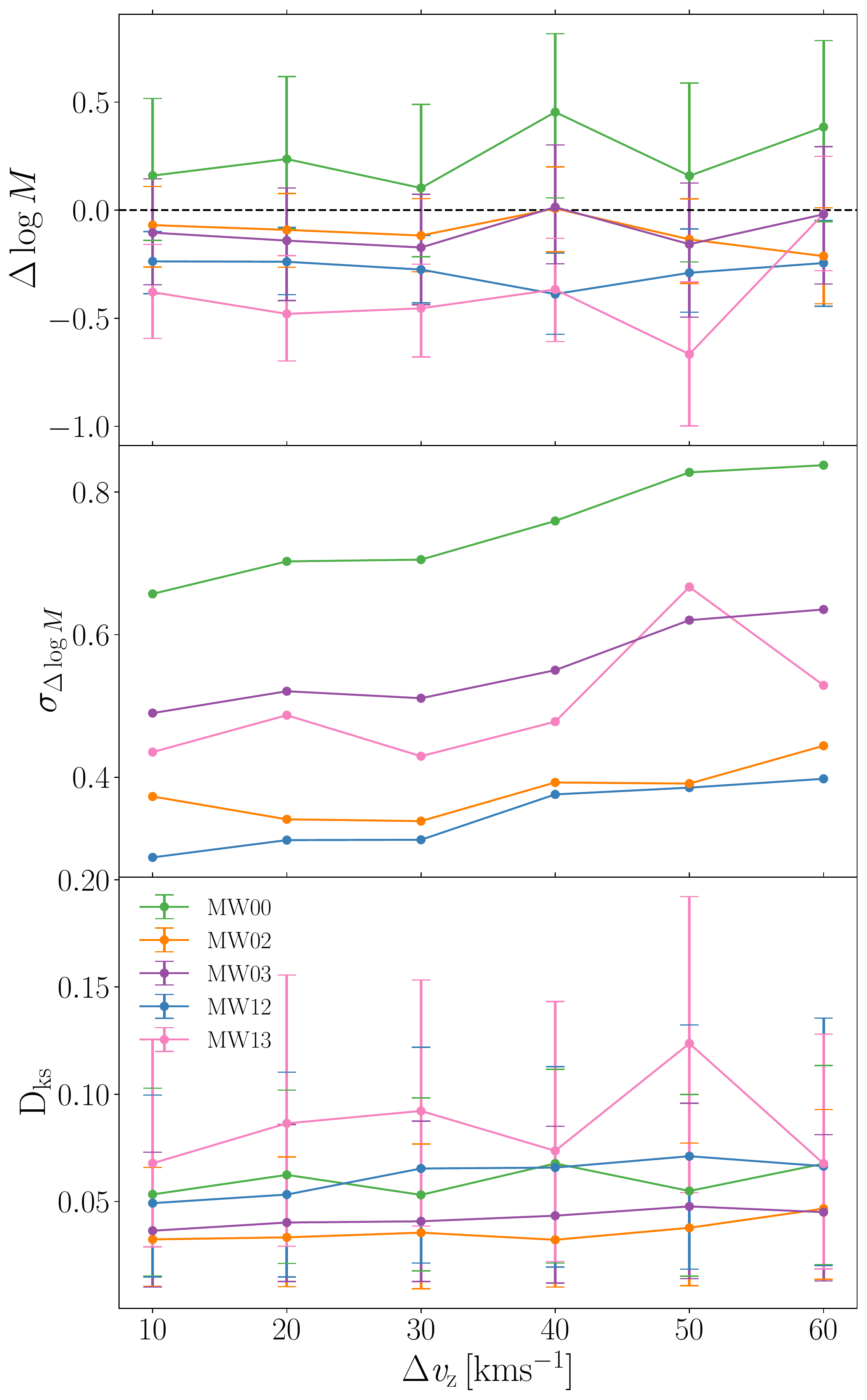}
    \caption{Estimators of the performance of the JAM model as in Fig. \ref{fig:ks_mass_agetest_newmass.pdf}, as a function of the line-of-sight velocity error. Here we show five galaxies: MW00, MW02, MW03, MW12 and MW13 as indicated by the legend.}
    \label{fig:ks_mass_error.pdf}
\end{figure}

So far, our analysis has been performed using a 10 $\kms$ line-of-sight velocity error. This, although achievable with surveys such as the Fornax 3D survey, is not always the error reached for all GCs (e.g. \citealt{Fahrion2020}). We therefore investigate the impact of increasing the error to 60 $\kms$ in increments of 10 $\kms$. For this, we take a similar approach to the previous section: we take the same four galaxies and rerun the JAM model using line-of-sight velocity errors $\Delta v_{\mathrm{z}} = 20, 30, 40, 50$ and $60\kms$. We then carry out our analysis in the same way as before for each of the runs.\par

The top panel of Fig. \ref{fig:ks_mass_error.pdf} shows the mass recovery. Similarly to Fig. \ref{fig:ks_mass_numberTest_GCnumber.pdf} there is no systematic variation of $\logm$ with increasing $\Delta v_{\mathrm{z}}$ but again there is a consistent offset, indicating that the model consistently over or underestimates the mass, regardless of $\Delta v_{\mathrm{z}}$ value. The $1 \sigma$ spread in the recovered mass ($\sigma_{\logm}$) tells a different story: it increases as $\Delta v_{\mathrm{z}}$ increases. It also seems to be reasonably stable until $\Delta v_{\mathrm{z}} = 30$ km/s and then the increase steepens towards $\Delta v_{\mathrm{z}} = 60$ km/s. However, there is an offset between the galaxies, with MW02 and MW12 (orange and blue points) always at lower $\sigma_{\logm}$ than MW03 and MW13 (purple and pink points). These galaxies always have lower $\sigma_{\logm}$ than MW00 (green points). Again, this is due to the number of GCs in each of these galaxies. From Table \ref{table:GCnumber} we know that MW02 and MW12 have the most GCs with $N_{\mathrm{GCs}} \approx 400$. MW03 has half this number with $N_{\mathrm{GCs}} \approx 200$ and MW13 and MW00 have a quarter with $N_{\mathrm{GCs}} \approx 100$. Therefore, although the velocity error does of course play a part in the recovery of the mass profile, it is the number of GCs that is the most important factor for the mass recovery. 

\subsection{Number of GCs and velocity error combined\label{6.3}}
\begin{figure}
	\includegraphics[width=\linewidth]{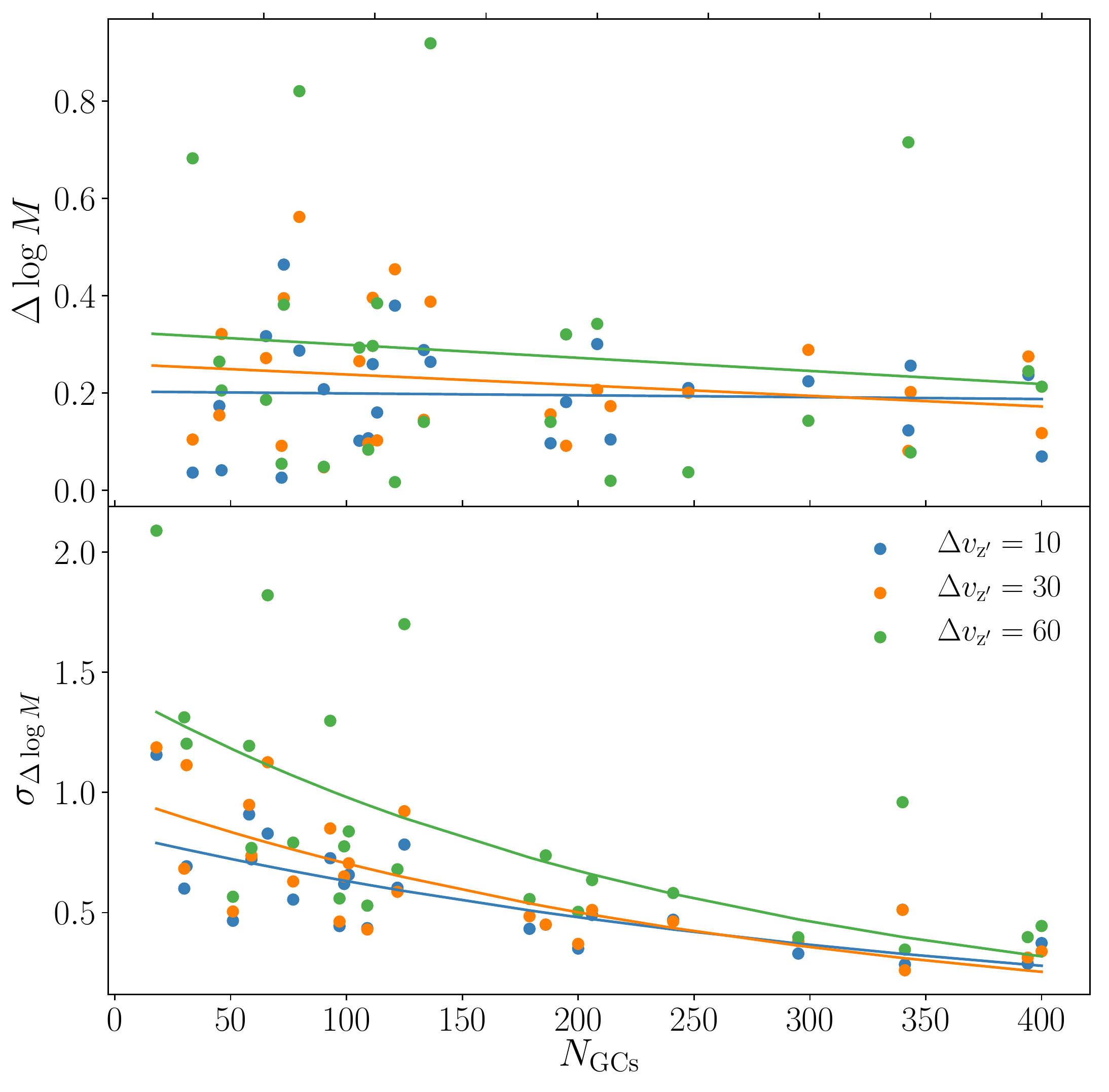}
    \caption{The accuracy and precision of the mass enclosed at the maximum GC radius for all 25 galaxies plotted as a function of the number of GCs in that galaxy. Each colour represents the results for three different line of sight velocity errors.}
    \label{fig:accuracy_precision.pdf}
\end{figure}
The previous two sections have shown that it is often a combination of factors that determine whether the JAM model is a good fit or not. Therefore, we now combine the effects of the number of GCs and the line-of-sight velocity error in Fig. \ref{fig:accuracy_precision.pdf}, where we show the mass enclosed recovery for the 25 galaxies as a function of the number of GCs in the galaxy and with three different velocity errors ($v_{\mathrm{z}}$= 10, 30, and 60$\kms$). The panels show the accuracy (i.e. how far from the true value we are, top panel) and the precision (i.e. the size of the error bar, bottom panel) of the enclosed DM mass at the maximum GC radius for each galaxy.  In the top panel we do not see an obvious trend between the accuracy and number of GCs, therefore we fit a linear model to each of the three $\Delta v_{\mathrm{z}}$ groups. The lines confirm that there is little to no improvement in the accuracy with increasing number of GCs. However, particularly for the galaxies with small numbers of GCs, the accuracy improves with smaller $\Delta v_{\mathrm{z}}$.\par

In the bottom panel of Fig. \ref{fig:accuracy_precision.pdf}, we show that the precision improves, as the number of GCs increases. We therefore fit each $\Delta v_{\mathrm{z}}$ group with an exponential 
\begin{equation}
\sigma_{\logm} = a e^{-N_{\mathrm{GCs}}/b},
\end{equation}
 
where $a$ and $b$ are free parameters.  All three curves show decreasing values of precision (therefore smaller error bars and better constraints on the mass enclosed) as the number of GCs increases. The exponential fits are: 
\begin{equation}
\sigma_{\logm} = 0.83 e^{-N_{\mathrm{GCs}}/367},
\end{equation}
\begin{equation}
\sigma_{\logm} = 0.99 e^{-N_{\mathrm{GCs}}/293},
\end{equation}
\begin{equation}
\sigma_{\logm} = 1.43 e^{-N_{\mathrm{GCs}}/266},
\end{equation}
for the $\Delta v_{\mathrm{z}}=10,30,60$ $\kms$ velocity errors respectively. \par

The $\Delta v_{\mathrm{z}}=60\kms$ (green) points are always higher than the $\Delta v_{\mathrm{z}}=30\kms$ (orange) points, which are always higher or the same as the $\Delta v_{\mathrm{z}}=10\kms$ points. This is true for the exponential fits and for each individual galaxy, meaning that the JAM models with the lower line-of-sight velocity errors constrain the mass better than the models with higher velocity error, as is expected. We also note that the blue and the orange points ($\Delta v_{\mathrm{z}}$ = 10 and 30 $\kms$ respectively) show less scatter around their exponential fit than the green ($\Delta v_{\mathrm{z}}$ = 60 $\kms$) points do.  What is particularly interesting about this panel is that the curves converge at the highest number of GCs, showing that as long as you have a high enough number of GCs it does not necessarily matter what the line-of-sight velocity error is. However, if you have a smaller number of GCs it will be of great benefit to reduce the line-of-sight velocity errors to obtain the best constraints on the mass enclosed. Finally, there is already a significant improvement in reducing the errors from 60 $\kms$ to 30 $\kms$ and there isn't a huge amount to be gained by improving the precision to 10 $\kms$. But of course this depends on the exact system being modelled, 10 km/s is around one tenth of the velocity dispersion of these systems. But for a galaxy cluster a higher error may suffice and for a lower mass galaxy higher precision measurements may be needed.

\section{Dependence on additional galaxy properties\label{7}}
There are many properties of a galaxy that could affect the performance of dynamical modelling. Therefore, we also carry out a similar analysis to \citet{Kruijssen2019} and search for statistical correlations between all galaxy assembly and formation properties with $\logm$, $\sigma_{\logm}$, $\mathrm{D_{KS}}$ and $\sigma_{\mathrm{D_{KS}}}$. We do this by calculating the Spearman rank-order correlation r and p values. We do not find significant correlations that are surprising or cannot be explained through a secondary corrrelation. For example, the number of mergers with a mass ratio < 1:100, the number of leaves in the merger tree and the number of branches in the merger tree all anti-correlate somewhat with $\sigma_{\logm}$, with an increase in the number of tiny mergers, leaves or branches resulting in a better constraint on the mass. All of these properties relate to the number of minor mergers, and \citet{Kruijssen2019} showed the number of minor mergers correlates with the number of GCs, therefore we interpret it as follows: the mass of a galaxy is better constrained when it has more minor mergers because this means that the galaxy has more GCs. We also find that $\sigma_{\logm}$ and $\sigma_{\mathrm{D_{KS}}}$ all correlate with the virial mass ($M_{200}$), the virial radius ($R_{200}$) and the maximum velocity ($V_{max}$), where an increase in the galaxy property means a better constraint on the mass and the radial profile of the DM halo. Again this can be explained in terms of the number of GCs because more massive haloes host more GCs and therefore the larger these particular galaxy properties are, the better constrained the properties of the DM halo become. \par

There are no correlations between how well the DM mass and profile is constrained and any other galaxy property. This means that the time of the last major merger does not have an effect on the JAM model. This is somewhat surprising since the JAM model assumes that the tracers and the potential are in equilibrium and a major merger would violate this assumption. The 25 galaxies span a broad range of lookback times for a major merger, from never experiencing one to having just undergone one. The lack of correlation could be explained by the relaxation time of the galaxy being relatively short- the galaxy returns to equilibrium within a short timescale after the final coalescence of the two progenitors.

\section{Summary and Conclusions\label{8}}
This work uses the E-MOSAICS suite of 25 zoom-in simulations of Milky Way mass galaxies to understand the extent to which different galaxy and GC properties, and data quality properties affect the outcomes of the axisymmetric Jeans model JAM \citep{Cappellari2008} using the approach for discrete tracers by \citet{Watkins2013}. This formalism of a Jeans model takes Multi Gaussian Expansion fits to the tracer population and the potential of the baryonic components as inputs and obtains a fit to the DM density profile within the radial range probed by the tracers. In our case the tracer population is the GCs and the potential of the baryons is obtained by fitting a Multi Gaussian Expansion to the stellar mass distribution. The DM component is parametrised using a gNFW profile. The gNFW profile has three free parameters and the JAM model adds one more in the form of the cylindrical velocity anisotropy ($\beta$), we therefore are left with a model with four free parameters. We explore this parameter space using an MCMC analysis.\par

We discuss in section \ref{4.1} and show in Figs. \ref{fig:Gal002_corner_2kpc.pdf} and \ref{fig:Gal004_corner_2kpc.pdf} that there are degeneracies between the DM halo parameters, even when fitting directly to the DM particle distribution from the simulation. Therefore we choose not to compare the recovery of the free parameters directly to the simulation DM profile fits. Instead we quantify how well the JAM model is performing with two quantities that probe the mass recovery and one quantity that probes the recovery of the DM radial profile. The enclosed DM mass recovery is quantified by the logarithmic difference between the DM mass within the maximum GC radius recovered by JAM and the truth from the simulation ($\logm$) and the $1 \sigma$ spread in this log difference from the spread in the posterior distributions from the JAM model ($\sigma_{\logm}$). The radial profile recovery is quantified using a KS test where we we calculate the maximum difference between two cumulative mass distributions ($\mathrm{D_{KS}}$). \par

Although all dynamical models perform well, some perform much better than others, as shown by our three diagnostics. We find that there is no effect on the DM profile recovery with the concentration or maximum radius of the GCs, but there is some effect due to the sphericity of the GC system. We fit the GC spatial distribution with a spherical MGE but it is clear from Fig. \ref{fig:ks_mass_sphericity.pdf} that most of the GC systems are not spherical.  More spherical GC systems result in a better constraint on the mass and DM profile. However, there is also another effect at play. The least spherical GC systems are also those with the fewest number of GCs and at a given sphericity the galaxies with the most GCs are also those with the best constraints. \par

Therefore, we also investigate how the number of GCs in the galaxy affects the recovery of the enclosed DM mass and radial distribution. Figs. \ref{fig:Gal002_mass_numberTest.pdf}-\ref{fig:ks_mass_numGCs.pdf} show that there is a strong dependency of the presicion of the JAM model on the number of GCs. The number of GCs has no effect on $\logm$ but a strong effect on $\sigma_{\logm}$ and also impacts our ability to constrain the DM halo profile. The dependency is exponential, where galaxies with fewer than 150 GCs show significant scatter in their mass and radial distribution recovery, but galaxies with more than 150 GCs consistently constrain the mass and distribution well. \par

We also investigate the impact of the data quality on the performance of the JAM model. This is done through increasing the line-of-sight velocity error. Fig.\ref{fig:ks_mass_error.pdf} shows that increasing the line-of-sight velocity error from 10 $\kms$ to 60 $\kms$  mainly has an impact on $\sigma_{\logm}$: the larger the error, the less well constrained the mass is. It also has a small effect on how well we constrain the DM halo radial profile. However, Fig. \ref{fig:ks_mass_error.pdf} also shows that the number of GCs plays a role in the recovery of the enclosed DM mass. Therefore, we combine the number of GCs and the velocity error in Fig. \ref{fig:accuracy_precision.pdf} and find that $\logm$ does not change with number of GCs, but becomes slightly worse when increasing $\Delta v_{\mathrm{z}}$. While $\sigma_{\logm}$ decreases exponentially with increasing number of GCs. When increasing $\Delta v_{\mathrm{z}}$ from 10 to 30 to 60 km/s there is a large difference in the recovery of the mass at low GC number, however, when large enough GC numbers are used the line-of-sight velocity error no longer plays a significant role in the recovery of the DM mass. We would therefore advise that if GC numbers are low in a spiral galaxy it is important for data to be of high quality.\par

The JAM model performs well for all of our 25 simulated galaxies. This is a promising result since the E-MOSAICS galaxies probe a wide range of formation and assembly histories and the JAM model can successfully deal with all of these. It also shows that GCs are effective as tracers in dynamical models and therefore it is possible to construct dynamical models of galaxies at higher redshift where stellar kinematics become problematic at large radii, but GCs remain bright tracers of the outer halo. The limitations of the JAM model come with the need to set a pre-defined geometry for the potential and - perhaps more importantly for a GC system - to assume a fixed orientation of the velocity ellipsoid. Some of these limitations can be alleviated by using more sophisticated dynamical modelling methods, e.g. higher order Jeans equations, distribution function based methods, and Schwarzchild orbit-based modelling techniques. Schwarzschild models, for example, allow the most flexibility in the geometry of the potential and the tracer distribution function enabling for a more rigorous dynamical description of the galaxies. It is unclear, however, how much data would be needed to well constrain these more sophisticated models which come with more free parameters. For our problem, where some galaxy halos are sampled with as few as 18 GC tracers, we decided to use simple axisymmetric Jeans models. For galaxies with more abundant data, revisiting this problem with more sophisticated techniques is highly desirable in the future. \par

\section*{Acknowledgements}
We thank the referee for their careful report that improved the clarity of this paper. We thank Dr Adrien Thob for providing the sphericity calculations for our GC populations. PJ and GvdV acknowledge funding from the European Research Council (ERC) under the European Union's Horizon 2020 research and innovation programme under grant agreement No 724857 (Consolidator Grant ArcheoDyn). JP and NB gratefully acknowledge funding from a European Research Council consolidator grant (ERC-CoG-646928-Multi-Pop). JMDK gratefully acknowledges funding from the German Research Foundation (DFG) in the form of an Emmy Noether Research Group (grant number KR4801/1-1). JMDK and MRC gratefully acknowledge funding from the European Research Council (ERC) under the European Unions Horizon 2020 research and innovation programme via the ERC Starting Grant MUSTANG (grant number 714907). MRC is supported by a Fellowship from the International Max Planck Research School for Astronomy and Cosmic Physics at the University of Heidelberg (IMPRS-HD). NB and RAC are Royal Society University Research Fellows. This work used the DiRAC Data Centric system at Durham University, operated by the Institute for Computational Cosmology on behalf of the STFC DiRAC HPC Facility (www.dirac.ac.uk). This equipment was funded by BIS National E-infrastructure capital grant ST/K00042X/1, STFC capital grants ST/H008519/1 and ST/K00087X/1, STFC DiRAC Operations grant ST/K003267/1 and Durham University. DiRAC is part of the National E-Infrastructure. The study also made use of high performance computing facilities at Liverpool John Moores University, partly funded by the Royal Society and LJMU's Faculty of Engineering and Technology.

\section*{Data Availability}
The data underlying this article will be shared on reasonable request to the corresponding author.

%%%%%%%%%%%%%%%%%%%%%%%%%%%%%%%%%%%%%%%%%%%%%%%%%%



\bibliographystyle{mnras}
\bibliography{bibliography.bib} % if your bibtex file is called example.bib

% Don't change these lines
\bsp	% typesetting comment
\label{lastpage}
\end{document}